\documentclass[%
 reprint,
 amsmath,amssymb,
prx,
]{revtex4-2}

\usepackage[T1]{fontenc}
\usepackage{amsmath,amssymb}
\usepackage{graphicx}% Include figure files
\usepackage{dcolumn}% Align table columns on decimal point
\usepackage{tabularray}
\usepackage{bm}% bold math
\usepackage{bbm}
\usepackage{blkarray}
\usepackage{caption}
\usepackage[colorlinks=true,citecolor=blue]{hyperref}
\usepackage[newcommands]{ragged2e} % This package and \justifying is needed for the caption of figures to be justified.
\usepackage{subcaption}

\usepackage{tikz}
\usetikzlibrary{quantikz}

\begin{document}

% \title{High-Distance Clifford-Deformed Local Fermionic Encodings}
\title{Low-Weight High-Distance Error Correcting Fermionic Encodings}
% \title{Clifford-deformed Fermionic Mappings with High Minimum Distance Stabilizer Codes}

% Force line breaks with \\

\author{Fedor \v{S}imkovic IV}
    \email{fedor.simkovic@meetiqm.com}
    \affiliation{IQM, Georg-Brauchle-Ring 23-25, 80992 Munich, Germany}
\author{Martin Leib}
    \affiliation{IQM, Georg-Brauchle-Ring 23-25, 80992 Munich, Germany}
    % \email{martin.leib@meetiqm.com}
\author{Francisco Revson F. Pereira}
    \affiliation{IQM, Georg-Brauchle-Ring 23-25, 80992 Munich, Germany}
    % \email{francisco.revson@meetiqm.com}
\date{\today}

\begin{abstract}
We perform an extended numerical search for practical fermion-to-qubit encodings with error correcting properties. Ideally, encodings should strike a balance between a number of the seemingly incompatible attributes, such as having a high minimum distance, low-weight fermionic logical operators, a small qubit to fermionic mode ratio and a simple qubit connectivity graph including ancilla qubits for the measurement of stabilizers. Our strategy consists of a three-step procedure in which we: first generate encodings with code distances up to  $d\leq4$ by a brute-force enumeration technique; subsequently, we use these encodings as starting points and apply Clifford deformations to them which allows us to identify higher-distance codes with $d\leq7$; finally, we optimize the hardware connectivity graphs of resulting encodings in terms of the graph thickness and the number of connections per qubit. We report multiple promising high-distance encodings which significantly improve the weights of stabilizers and logical operators compared to previously reported alternatives.

\end{abstract}

\maketitle

%\tableofcontents

\section{Introduction}
\label{sec:Introduction}
The numerous advances of recent years are bringing the prospect of useful quantum computation ever closer to reality. Multiple fault-tolerant algorithms with speedup guarantees have been proposed~\cite{Dalzell2023}, however the computational overhead in terms of qubit numbers and circuit depths is rendering them out of reach for current generations of quantum processors, as well as any future ones for years to come. In contrast, current NISQ devices have not yet been able to provably demonstrate quantum advantage despite tremendous efforts in improving both hardware and algorithms. The fundamental reason for this is that such noisy devices are limited to only relatively short circuit depths, which allows for classical methods to retain their computational superiority. The central question is whether it is possible to bridge the gap between the computational eras of NISQ and fault tolerant quantum computing by introducing a trade-off between fidelities and circuit depth with other, more readily available resources, such as the number of qubits and computational time in order to improve existing quantum computations.

On the one hand, various error mitigation strategies have been proposed for this purpose, which attempt to improve calculations by either extrapolating from or averaging over noisy data at the cost of an exponentially scaling overhead in the number of repetitions~\cite{cai2023quantum}. On the other hand, preliminary forms of (partially) error-corrected algorithms have been proposed with the idea of correcting at least low-weight and high-probability errors at the expense of utilizing additional qubit resources~\cite{Haah2017,setia2019superfast,Bausch2020,Tantivasadakarn2020,chen2022error}. In this paper, we will focus our attention on the latter of these two approaches.

One of the most promising classes of models for simulation on early quantum devices are fermionic quantum systems. These models are in many cases extremely challenging for classical algorithms due to their large Hilbert spaces and infamous fermionic sign problem, which arises as a consequence of the Pauli exclusion principle. As a result, even deceptively simple problems, such as the paradigmatic Fermi-Hubbard model~\cite{arovas2022hubbard, qin2022hubbard}, have not yet been fully understood in their most interesting and simultaneously challenging regimes.

In order to port a fermionic Hamiltonian to a quantum computer, it is necessary to encode the underlying indistinguishable fermionic modes in terms of distinguishable spins. In one dimension this encoding is trivially performed using the Jordan-Wigner transformation (JWT)~\cite{Jordan1928}, which maps every fermionic mode to a single qubit. In higher dimensions, however, the JWT leads to increasingly non-local interactions as the system size grows. For this reason, a number of fermion-to-qubit encodings have been developed, which utilize additional ancillary qubits in order to fix fermionic commutation relations locally~\cite{Bravyi2017,Havlicek2017,Steudtner2018,jiang2019majorana,setia2019superfast,Bausch2020,Tantivasadakarn2020,chen2022error,Harrison2022,Kirby2022,chien2022optimizing,Chiew2023,algaba2023low}. As a side effect of the extended Hilbert space, stabilizers materialize, which commute with all logical operators of the underlying Hamiltonian. These, in turn, can be used to identify and/or correct low-weight Pauli errors which may appear during circuit execution. 

Designing performant fermionic encodings with quantum error correction in mind initially seems to pursue a set of contradicting goals. From the simulation point of view one is interested in keeping the weights of all logical operators present in the Hamiltonian as low as possible in order to reduce the hardware requirements in terms of the total gate count and circuit depth. At the same time, the minimum distance, a measure of the maximum weight of errors that can be identified or corrected, is limited by the size of the smallest logical operator. Additionally, it is important to keep stabilizer operators geometrically local and low-weight in order to minimize the overhead of mid-circuit error-correcting cycles and maximize parallelism. Finding encodings which simultaneously cater to both of these paradigms is a highly non-trivial classical optimisation problem. Nevertheless, a number of recent studies pursuing brute-force methods have been able to identify promising candidate encodings with distances in the range of $2\leq d \leq7$ \cite{setia2019superfast, jiang2019majorana, chen2022error,chien2022optimizing}. Another aspect, which is important in terms of the realisations of such fermionic quantum error correcting encodings (FQECE) on realistic quantum hardware is the geometric locality and connectivity of qubits. In literature, it is often assumed that qubits can have arbitrary (all-to-all) connectivity and that additional syndrome qubits (including their connectivity), which are needed for the measurement of stabilizers, are freely available within the quantum hardware. Another limitation of many current quantum architectures is that they do not allow for any non-planar couplings within the connectivity graphs, which adds to the difficulty of identifying suitable FQECEs.

In this work, we perform an extensive search for efficient FQECEs which enable the simulation of Fermi-Hubbard-type interactions between nearest- and next-nearest-neighbors (NN and NNN respectively). We first implement a brute-force search algorithm, which identifies suitable encodings by iterating through all \emph{reasonable} combinations of logical operators. This procedure generates FQECEs with distances $d \leq 4$, which are optimized in terms of the weights of both logical operators and stabilizers, and also reproduces a number of known fermion-to-qubit encodings from literature \cite{verstraete2005mapping, derby2021compact,algaba2023low}. We then use these FQECEs as starting points for a second algorithm which transforms between different encodings by means of Clifford deformations and identify additional state-of-the-art FQECEs up to distance $d \leq 7$. In both of these algorithms we only impose the size of a translationally invariant unit cell of an encoding, which can consist of up to six qubits. In order to adhere to potential hardware restrictions, we numerically optimize the maximum connectivity of qubits for various FQECEs whilst keeping the thickness of the qubit connectivity graph low. Finally, we propose co-design strategies in terms of connectivities between qubits and stabilizers. For higher-distance encodings, these can involve either bi-planar hardware connectivity graphs or, alternatively, a limited number of SWAP operations.

The paper is structured as follows: In Section~\ref{sec:Preliminaries}, we introduce the formalism behind local fermion-to-qubit encodings, quantum stabilizer codes and Clifford deformations thereof. In Section~\ref{sec:Generation}, we describe  our implementations of two FQECE-search algorithms as well as the restrictions imposed on the classes of FQECEs of interest to us. We present the overall results of our extensive search for optimal FQECEs and compare them to previous studies in Section~\ref{sec:Results}. In Section~\ref{sec:Codesign} we discuss in detail the hardware requirements and potential for co-design implementations of the FQECEs we have identified. We provide an outlook and concluding remarks on the usefulness of FQECEs in the beyond-NISQ era in Section~\ref{sec:Conclusions}.

\section{Preliminaries}
\label{sec:Preliminaries}
\subsection{Fermion-to-Qubit Encodings}

The simulation of fermionic systems on quantum computers is implemented by encoding relevant fermionic operators into strings of Pauli operators. This encoding allows us to represent a fermionic Hamiltonian as a linear combination of tensor products of Pauli operators. This new representation of fermionic systems can be used by a quantum computer to, e.g., simulate time evolution, which can be a subroutine of many other quantum algorithms. The Jordan-Wigner transformation (JWT) is the best known fermion-to-qubit encoding. Consider the fermionic creation and annihilation operators obeying the following commutation relations:
\begin{equation}
    \{c_i^\dagger, c_j^\dagger\}=\{c_i,c_j\}=0, \quad \{c_i^\dagger,c_j\}=\delta_{ij}.
\end{equation}Then, the JWT represents these operators $c_i^\dagger$ and $c_i$ as
\begin{eqnarray}
    c_i^\dagger &\rightarrow& Z_1\cdots Z_{i-1}\sigma_i^+,\\
    c_i &\rightarrow& Z_1\cdots Z_{i-1}\sigma_i^-,
\end{eqnarray}where $\sigma_i^+ \equiv  (X_i - iY_i)/2$ and $\sigma_i^- \equiv  (X_i + iY_i)/2$. One of the disadvantages of using the JWT is that even if a fermionic operator acts on a fixed number of modes, the corresponding spin representation can act on up to $O(m)$ qubits, where $m$ is the total number of modes in the fermionic system.

To overcome the locality issue present in the JWT, several \emph{local encodings} with the property of representing geometrically local fermionic operators as geometrically local strings of Pauli operators have been proposed~\cite{ball2005fermions, verstraete2005mapping, steudtner2017lowering, chen2018exact, derby2021compact, algaba2023low}. Rather than directly encoding fermionic creation and annihilation operators into Pauli operators, one instead goes through the intermediate representation of fermionic edge and vertex operators, defined as
\begin{equation}
    E_{ij} \equiv  -i\gamma_i\gamma_j, \quad V_i \equiv  -i\gamma_i \bar{\gamma_i},
\end{equation}respectively, where $\gamma_j \equiv  c_j^{\dag} + c_j$ and $\overline{\gamma}_j \equiv  i(c_j^{\dag} - c_j)$ are Majorana operators. From the commutation relations of the fermionic creation and annihilation operators one can derive the corresponding commutation relations between vertex and edge operators:
\begin{equation}
    [E_{ij},V_l] = [V_i,V_j] = [E_{ij},E_{ln}] = 0,
\end{equation}for all $i\ne j \ne l \ne n$, and 
\begin{eqnarray}
    \{E_{ij}, E_{jk}\} = \{E_{ij},V_i\} = 0.
\end{eqnarray}
In words, if one constructs a graph out of edge and vertex operators, then all edges incident on a given vertex anticommute with the vertex and between themselves and in all other cases operators commute. One useful feature is that one can define an edge operator between any two vertices as long as an edge path between them exists by using the composite edge identity: 
\begin{equation} \label{eq:composite}
E_{ik}=iE_{ij}E_{jk}.
\end{equation}
Additionally, it is easy to show that every closed loop of edges forms a stabilizer operator, which commutes with all logical edge and vertex operators. The FQECE is then defined in the joint $+1$ subspace of all stabilizers. More precisely,
\begin{equation}\label{eq:stabilizer} i^{(|p|-1)}\prod_j^{|p|-1} E_{p_j,p_{j+1}} = \mathbbm{1},
\end{equation}where $p=\{p_1, p_2, \ldots, p_n\}$ is the set of sites. Within a given fermion-to-qubit encoding, we can then associate Pauli strings to stabilizer closed loops of edges. This establishes the connection between encodings and stabilizer codes.

\subsection{Majorana Monomial Representation}

The representation of fermionic operators can be made clearer by the use of the group of Majorana monomials, which has edge and vertex operators as elements. Let $\gamma_j$ and $\overline{\gamma}_j$ be the Majorana operators, and $\mathbb{F}_2$ be the binary field, then the group of Majorana monomials is given by
\begin{eqnarray}
\nonumber \mathcal{M} \equiv  &\left\lbrace M(\mathbf{b})\equiv \prod_{j=1}^{m} \gamma_j^{b_j} \overline\gamma_j^{b_{j+m}} \;\middle \vert \; \mathbf{b} \in \mathbb{F}_2^{2m}  \right\rbrace\times\\
 & \{\pm 1, \pm i \}.
\label{majorana_group}
\end{eqnarray}
In this paper, we are interested in $F\leq\mathcal{M}$, which represents the underlying physical system. Since a fermion-to-qubit encoding consists of a representation of the Majorana monomials preserving the commutation relations between observables, we have that such encodings must be an algebra isomorphism. Consider that the subgroup of interest, $F$, is generated by a set $\mathcal{F}=\{f_1, \ldots, f_{|\mathcal{F}|}\}$. We then only need to describe the fermion-to-qubit encoding over the elements of the (finite) basis $\mathcal{F}$. Consider the finitely generated group:
\begin{equation}
\Gamma= \left\lbrace \Gamma(\mathbf{b})\equiv \prod_{j=1}^{\vert \mathcal{F}\vert} f_j^{b_j} \;\middle \vert \; \mathbf{b} \in \mathbb{F}_2^{\vert \mathcal{F} \vert}  \right\rbrace \times \{\pm 1, \pm i \}.
\label{eq:gamma_group}
\end{equation}
Let $\tau: \Gamma \rightarrow F$ be the group homomorphism described by the action on the group generating elements $\tau(f_i \in \Gamma )\equiv  f_i \in F$. In particular, one can see that $F \simeq \Gamma / \ker(\tau)$. The fermion-to-qubit encoding is denoted by $\sigma$ and is described as an encoding $\sigma: F \rightarrow \mathcal{P}$, where 
\begin{eqnarray}
\mathcal{P} \equiv  & \left\lbrace P(\mathbf{b})\equiv \prod_{j=1}^{n} X_j^{b_j}Z_j^{b_{j+n}} \;\middle \vert \; \mathbf{b} \in \mathbb{F}_2^{2n}  \right\rbrace \times\nonumber\\
&\{\pm 1, \pm i \}
\label{pauli_group}
\end{eqnarray}is the Pauli group over $n$ qubits. For $\sigma$ to be an algebra isomorphism, it must preserve the group commutation relations:
\begin{equation}
\forall \; f_i, f_j \in \mathcal{F}: \;[f_i,f_j]\equiv f_i^{-1}f_j^{-1}f_i f_j=\pm1
\label{commutation}
\end{equation}
and inverse/hermiticity relations:
\begin{equation} \label{inverse}
 \forall f \in \mathcal{F} :\;  f^{-1}=f^\dagger = \pm f 
\end{equation}among all elements of $\mathcal{F}$. It was shown in Ref.~\cite{chien2023simulating} that if the encoding $\sigma$ exists, then one can identify a (stabilizer) abelian subgroup $\mathcal{S} \equiv  \sigma(\text{ker}(\tau))$ on $\mathcal{P}$ 
and a superselection subgroup $G \equiv  \tau(\text{ker}(\sigma))$ on $F$. By fixing the superselection group G to be $+1$, one can extend the encoding $\sigma$ to an algebra isomorphism on the group algebra $\mathbb{C}[F/G]$ by projecting into the code space $V$ stabilized by $\mathcal{S}$.

\subsection{Quantum Stabilizer Codes} 

A stabilizer code $\mathcal{Q}$ is a subspace of an $n$-qubit system stabilized by the elements of an abelian subgroup $\mathcal{S}$ of the operator group acting over $n$ qubits. For simplicity, we are going to assume that $\mathcal{S}$ belongs to the Pauli group over $n$ qubits.

The commutation relations between elements in the Pauli group can be described by a symplectic form over a binary vector space. 
Let $A$ be a Pauli operator over $n$ qubits. Up to a phase, we can write $A=X(\mathbf{a})Z(\mathbf{b})$, with $X(\mathbf{a})=X^{a_1}\otimes\cdots\otimes X^{a_n}$ and $Z(\mathbf{b})=Z^{b_{1}}\otimes\cdots\otimes Z^{b_{n}}$ where $\mathbf{a}=(a_1,\ldots,a_n),\mathbf{b}=(b_1,\ldots,b_n)\in\mathbb{F}_2^{n}$. Using this description, we can describe the commutation relation between two operators $E=X(\mathbf{a})Z(\mathbf{b})$ and $E'=X(\mathbf{a}')Z(\mathbf{b}')$ by:
\begin{equation}
[E,E'] = (-1)^{\omega_q((\mathbf{a}|\mathbf{b}),(\mathbf{a}'|\mathbf{b}'))},
\end{equation}where
\begin{align}
    \omega_q((\mathbf{a}|\mathbf{b}),(\mathbf{a}'|\mathbf{b}')) &= (\mathbf{a}|\mathbf{b})^T \Lambda_q (\mathbf{a}'|\mathbf{b}'), \nonumber \\
    \text{with}\quad \Lambda_q &= 
    \begin{pmatrix}
    0 & 1\\
    1 & 0
    \end{pmatrix} \otimes I_{n\times n},
     \label{eq:pauli_symplectic}
\end{align}
is a binary symplectic form. Since the stabilizer group is abelian, we have that $\omega_q((\mathbf{a}|\mathbf{b}),(\mathbf{a}'|\mathbf{b}'))=0$ for every pair $S_1=X(\mathbf{a})Z(\mathbf{b})$ and $S_2=X(\mathbf{a}')Z(\mathbf{b}')$ in $\mathcal{S}$.

The description of correctable and uncorrectable errors can also be given in terms of the binary representation. Let $C \equiv  \{(\mathbf{a}|\mathbf{b})\; \vert \; i^c X(\mathbf{a})Z(\mathbf{b}) \in \mathcal{S} \text{ for some }c\in\{0,1,2,3\}\}$. Then a code $C\subseteq\mathbb{F}_2^{2n}$ detects errors with weight less than
\begin{align}
& d_\text{min} \equiv \\ & \min\{\text{swt}((\mathbf{a}'|\mathbf{b}')) \; \vert\; \omega_q(((\mathbf{a}|\mathbf{b}),(\mathbf{a}'|\mathbf{b}'))) = 0, \forall (\mathbf{a}|\mathbf{b})\in C\}, \nonumber
\end{align}where $\text{swt}((\mathbf{a}|\mathbf{b})) = |\{k\colon (a_k,b_{k})\ne (0,0)\}|$. We call $d_\text{min}$ the minimum distance of the code. For some particular families of quantum error-correcting codes derived algebraically, one can compute the minimum distance from the properties of the classical error-correcting codes used in the construction; e.g., quantum error-correcting codes derived from the Calderbank-Shor-Steane (CSS) construction~\cite{Nielsen2010}, where one can relate the minimum distance of the quantum code to the minimum distance of the classical codes used in the construction. We can also compute the minimum distance from intrinsic properties of a quantum error-correcting code; e.g., 
one can compute the minimum distance of the surface code from the topological properties present in it~\cite{Fowler2012}. However, it is known that computing the minimum distance of classical and quantum error-correcting codes is an NP-hard problem~\cite{Vardy1997,Kapshikar2023}. The quantum error-correcting codes that we derive in this paper are obtained from brute-force search and Clifford deformations, making it hard to spot the underlying classical error-correcting code and related minimum distance. Therefore, we compute the minimum distance of the code exactly by generating all possible Pauli errors, where we gradually increase the operator weight. As the code distance corresponds to the minimal non-zero operator weight of an error that commutes with all stabilizer generators, we stop when this error is found. 

\subsection{Binary Representations of Fermionic Systems}

The classical complexity for describing arbitrary operators acting on a Hilbert space grows exponentially in the number qubits. Notwithstanding, one can consider a subgroup of the operator group that has a computationally feasible description via a binary vector \cite{Aaronson2004}. In particular, both groups of operators considered in this paper, namely the Majorana group $\mathcal{M}$ and the Pauli group $\mathcal{P}$, have this property. Furthermore, we can extend the action of these binary representations in order to guarantee that operator multiplication corresponds to vector addition, turning them, therefore, to isomorphisms.

Let $\tau$ and $\sigma$ be isomorphisms from the basis $\mathcal{F}$ to the group of
Majorana monomials and to the Pauli group, respectively. Suppose $\hat{\tau}$ and $\hat{\sigma}$ are projective matrix representations over the binary field of the respective isomorphisms. The matrices $\hat{\tau}$ and $\hat{\sigma}$ have the following structure: 
\begin{equation}
\label{eq:tau}
\hat{\tau} = 
\begin{blockarray}{ccccc}
&f_1 & f_2 & \cdots & f_{|\mathcal{F}|} \\
\begin{block}{c(cccc)}
\gamma_1 \phantom{,} & \tau_{1,1} & \tau_{1,2} & \cdots & \tau_{1,|\mathcal{F}|} \\
\gamma_2 \phantom{,} & \tau_{2,1} & \tau_{2,2} & \cdots & \tau_{2,|\mathcal{F}|} \\
\vdots \phantom{,} & \vdots & \vdots & \ddots & \vdots  \\
\gamma_{m} \phantom{,} & \tau_{m,1} & \tau_{m,2} & \cdots & \tau_{m,|\mathcal{F}|} \\
\bar{\gamma}_1 \phantom{,} & \tau_{m+1,1} & \tau_{m+1,2} & \cdots & \tau_{m+1,|\mathcal{F}|} \\
\bar{\gamma}_2 \phantom{,} & \tau_{m+2,1} & \tau_{m+2,2} & \cdots & \tau_{m+2,|\mathcal{F}|} \\
\vdots \phantom{,} & \vdots & \vdots & \ddots & \vdots \\
\bar{\gamma}_{m} \phantom{,} & \tau_{2m,1} & \tau_{2m,2} & \cdots & \tau_{2m,|\mathcal{F}|} \\
\end{block}
\end{blockarray} 
\end{equation}and
\begin{equation}
\label{eq:sigma}
\hat{\sigma} = 
\begin{blockarray}{ccccc}
&f_1 & f_2 & \cdots & f_{|\mathcal{F}|} \\
\begin{block}{c(cccc)}
Z_1 \phantom{,} & \sigma_{1,1} & \sigma_{1,2} &  \cdots & \sigma_{1,|\mathcal{F}|} \\
Z_2 \phantom{,} & \sigma_{2,1} & \sigma_{2,2} & \cdots & \sigma_{2,|\mathcal{F}|} \\
\vdots \phantom{,} & \vdots & \vdots & \ddots & \vdots  \\
Z_{n} \phantom{,} & \sigma_{n,1} & \sigma_{n,2} &  \cdots & \sigma_{n,|\mathcal{F}|}\\
X_1 \phantom{,} & \sigma_{n+1,1} & \sigma_{n+1,2} &  \cdots & \sigma_{n+1,|\mathcal{F}|} \\
X_2 \phantom{,} & \sigma_{n+2,1} & \sigma_{n+2,2} &  \cdots & \sigma_{n+2,|\mathcal{F}|} \\
\vdots \phantom{,} & \vdots & \vdots  & \ddots & \vdots \\
X_{n} \phantom{,} & \sigma_{2n,1} & \sigma_{2n,2} &  \cdots & \sigma_{2n,|\mathcal{F}|}\\
\end{block}
\end{blockarray}\hspace{0.2cm},
\end{equation}respectively, where $\hat{\tau}(f_i) = \delta_\tau(f_i)M(\hat{\tau}_i)$ and $\hat{\sigma}(f_i) = \delta_\sigma(f_i)P(\hat{\sigma}_i)$. The precise phases $\delta_\tau$ and $\delta_\sigma$ of the binary representations $\tau$ and $\sigma$, respectively, are not important for our purposes, and we refer the reader to Ref.~\cite{chien2022optimizing} for details on how to determine them. 

From Eqs. (\ref{eq:tau}) and (\ref{eq:sigma}), one can clearly see that the group commutation relations can be expressed in terms of matrix equations. Similarly to the Pauli group case, let $\tilde{\mathbf{a}},\tilde{\mathbf{b}}\in\mathbb{F}_2^{2n}$. Following the commutation relation from Eq. (\ref{commutation}), we have that any two Majorana operators $M(\tilde{\mathbf{a}}), M(\tilde{\mathbf{b}})$ obey
\begin{equation}
    [M(\tilde{\mathbf{a}}), M(\tilde{\mathbf{b}})] = (-1)^{\omega_f(\tilde{\mathbf{a}}, \tilde{\mathbf{b}})},
\end{equation}
where 
\begin{equation}\label{eq:fermi_symplectic}
   \omega_f(\tilde{\mathbf{a}}, \tilde{\mathbf{b}}) = \tilde{\mathbf{a}}^T \Lambda_f \tilde{\mathbf{b}}, \quad \Lambda_f = I + C_1
\end{equation}
where $I$ is the $2n\times 2n$ identity matrix and $C_1$ is the constant matrix with $1$ in every entry.

Now, we can describe the  group commutation relations when representing a fermionic system over the group of Majorana monomials and the Pauli group. Since both encodings describe the same fermionic system, we must have the same group commutation relations, which can be translated to the condition:
\begin{equation}
\hat{\tau}^\dag \Lambda_f \hat{\tau} = \hat{\sigma}^\dag \Lambda_q \hat{\sigma}.
\label{matrix_comm_cond}
\end{equation}

\subsection{Translational Invariance}

We have discussed the types of fermionic operators that we are interested in and shown that they can be described by matrices, where a careful description can resolve the locality issue present in some encodings, such as the JWT~\cite{Jordan1928}.
We focus in this paper on translationally invariant fermionic systems, which is a property we aim to preserve in the respective qubit representations thereof. For the purpose of this paper, translational invariance is of the form of a two-dimensional tessellation described by a lattice. As we are going to show below, encodings that preserve translational invariance can be described by multi-variable Laurent polynomials. The formalism that we follow here was first introduced by Haah~\cite{Haah2017} and previously adapted to the context of fermion-to-qubit encodings by Chien and Klassen~\cite{chien2022optimizing}.

A lattice is a collection of sites, modeled by the additive group $\mathbb{Z}_D$. Elements in $\mathbb{Z}_D$ can be described by the polynomial ring in $D$ variables $x_1,\ldots, x_D$,  multi-variable Laurent polynomials. Actions on this lattice are implemented by polynomial multiplication. In particular, translation along the $x_1$-direction is performed by multiplying the initial Laurent polynomial by $x_1$. The representation of Majorana monomials, Pauli operators, related groups, and encodings is given by the ring
\begin{multline}
\mathbb{F}_2[x_1,..,x_D]^{n \times m}\equiv  \\ \left\lbrace \sum_{\mathbf{k} \in \mathbb{Z}^D} a_{\mathbf{k}} \prod_{i=1}^D x_i^{k_i} \;\middle \vert \; a_{\mathbf{k}} \in \mathbb{F}_2^{n \times m} \right \rbrace,
\end{multline}
where $\mathbf{k} = (k_1,\ldots,k_D)$. We also consider conjugation of an element $a =  \sum_{\mathbf{k} \in \mathbb{Z}^D} a_{\mathbf{k}} \prod_{i=1}^D x_i^{k_i}$ as $a^\dagger =  \sum_{\mathbf{k} \in \mathbb{Z}^D} a_{\mathbf{k}}^{\mathsf{T}} \prod_{i=1}^D x_i^{-k_i}$ and the translation in the lattice by a vector $\mathbf{k}$ as $T_{\mathbf{k}} \equiv  \prod_{i=1}^D x_i^{k_i}$. Extending our formulation to Pauli operators $\mathcal{P}$, the finitely generated group $\Gamma$, and Majorana monomials $\mathcal{M}$, we have
\begin{align}
P(a)&\equiv  \prod_{\mathbf{k} \in \mathbb{Z}^D} \prod_{j=1}^{n_c} {X_{j,\mathbf{k}}^{a_\mathbf{k}[j]}} {Z_{j,\mathbf{k}}^{a_\mathbf{k}[j+n]}}, \\
\Gamma(a)&\equiv  \prod_{\mathbf{k} \in \mathbb{Z}^D} \prod_{j=1}^{J} {f_{j,\mathbf{k}}^{a_\mathbf{k}[j]}},  \\
M(a)&\equiv  \prod_{\mathbf{k} \in \mathbb{Z}^D} \prod_{j=1}^{m} {\gamma_{j,\mathbf{k}}^{a_\mathbf{k}[j]}}{\bar{\gamma}_{j,\mathbf{k}}^{a_\mathbf{k}[j+m]}},
\end{align}respectively, where $a= \sum_{\mathbf{k} \in \mathbb{Z}^D} a_{\mathbf{k}} T_{\mathbf{k}}$, the multi index $(j,\mathbf{k})$ indicates the $j$th qubit, Majorana monomials $f_j$ or Majorana mode $\gamma_j$, respectively, acting on a unit cell of the lattice translated from the origin by $\mathbf{k}$. Further, $a_{\mathbf{k}} [j]$ is the $j$th entry of $a_\mathbf{k}$. Notice that we are considering that each unit cell contains $n_c$ qubits, $J$ Majorana monomials or $m$ modes, respectively. 
% The products are always ordered in accordance with any preferred ordering on the vectors $\mathbf{k}$. 

The description of translational invariance via multi-variable Laurent polynomials can also be extended to the commutation relations described in Eqs.~(\ref{eq:pauli_symplectic}) and (\ref{eq:fermi_symplectic}). Let $\Lambda_Q$ and $\Lambda_F$ be the multi-variable Laurent polynomial descriptions of the commutation relations in Eqs.~(\ref{eq:pauli_symplectic}) and (\ref{eq:fermi_symplectic}), respectively. Then, we have that:
\begin{equation}
\Lambda_Q \equiv   \Lambda_q,
\end{equation}
and
\begin{equation}
\Lambda_F \equiv  I + \sum_{\mathbf{k} \in \mathbb{Z}^D} C_1 T_{\mathbf{k}}.
\end{equation}
Now, the commutation relations can be expressed by the following relation:
\begin{equation}
w_{Q/F}(a,b)\equiv  (a^\dagger \Lambda_{Q/F} b)_{\mathbf{k}},
\end{equation}
where the other terms contain information about commutation relations between relative translated version of $b$, 
\begin{equation} 
(a^\dagger \Lambda_{Q/F} b)_{\mathbf{k}} \equiv  (a^\dagger \Lambda_{Q/F} b T_{\mathbf{k}})_{\mathbf{0}}= w_{Q/F}(a,b T_{\mathbf{k}}).
\end{equation}

\subsection{Clifford Transformations}

We have shown a procedure to represent fermionic operators as strings of Pauli operators. From the underlying structure of the fermionic and Pauli operators, we were able to give a binary vector space description and establish algebraic relations between fermionic elements and their Pauli representation. 
In order to generate different representations of the same fermionic operators and simultaneously keep the algebraic relations between elements, we are going to employ the Clifford group as the mapping connecting two distinct representations of the same fermionic system in terms of different strings of Pauli operators. 
Since the elements in the Clifford group are unitary operators that map Pauli operators to Pauli operators, the symplectic form description is still valid for any output representation. Thus, the commutation relations of the input representation are preserved under the action of Clifford operators, whilst this action can be described using the same symplectic formalism.

Let $\mathcal{P}$ be the Pauli group over $n$ qubits defined in Eq.~(\ref{pauli_group}) and $U_{2^n}$ be the group of unitary matrices over $n$ qubits. The Clifford group is given by the group of unitary transformations that normalize the Pauli group
\begin{equation}
\mathcal{C} \equiv   \left\lbrace V\in U_{2^n} \;\middle \vert \; VPV^\dagger \in \mathcal{P} \text{ for any } P\in\mathcal{P}\right\rbrace.
\label{clifford_group}
\end{equation}The elements in the Clifford group are called Clifford gates. It is possible to show that the Clifford group is finite and finitely generated~\cite{Gottesman_1998, gottesman1998heisenberg}. In particular, we are going to use $H$, $S$, and CX gates as generators of the Clifford group. $H$ is the Hadamard gate and can be written as 
\begin{equation}
H = \frac{1}{\sqrt{2}}
\begin{pmatrix}
 1 & 1 \\
 1 & -1 \\
\end{pmatrix},
\end{equation}which maps $X$ gates to $Z$ gates and vice-versa; i.e., $H X H^\dagger = Z$ and $H Z H^\dagger = X$. The S gate, also called phase gate, has matrix representation
\begin{equation}
S = 
\begin{pmatrix}
 1 & 0 \\
 0 & e^{i\frac{\pi}{2}} \\
\end{pmatrix}.
\end{equation}
It is a Clifford gate since it maps $S X S^\dagger = Y$ and $S Z S^\dagger = Z$. Lastly, the CX gate is the two-qubit gate required to generate the Clifford group and given by 
\begin{equation}
\text{CX} = 
\begin{pmatrix}
 1 & 0 & 0 & 0 \\
 0 & 1 & 0 & 0 \\
 0 & 0 & 0 & 1 \\
 0 & 0 & 1 & 0 \\
\end{pmatrix}.
\end{equation}The action of the CX gate on the Pauli group over two qubits is described by the following relations
\begin{eqnarray}
    \text{CX}(X\otimes I)\text{CX}^\dagger &=& X\otimes X,\nonumber\\
    \text{CX}(Z\otimes I)\text{CX}^\dagger &=& Z\otimes I,\nonumber\\
    \text{CX}(I\otimes X)\text{CX}^\dagger &=& I\otimes X,\nonumber\\
    \text{CX}(I\otimes Z)\text{CX}^\dagger &=& Z\otimes Z.
\end{eqnarray}For the case when there is a lattice describing the connectivity between qubits, and therefore the possible two-qubit gates, only a subset of these transformations is needed. We describe our approach to picking the subset in Section~\ref{sec:clifford_search}.

\section{Generation of Fermionic Encodings}
\label{sec:Generation}
\subsection{Fermionic Hamiltonians}

In order to be able to compare the performance of different fermion-to-qubit encodings in terms of logical operator weights (which are related to the total number of quantum gates and circuit depth) we must pick a specific fermionic Hamiltonian of interest. For this purpose we consider the Fermi-Hubbard model (FHM) on the square lattice,  which is widely believed to be a promising candidate for early quantum advantage. The second-quantized FHM Hamiltonian is given by:
\begin{align}
\mathcal{H}_{\text{FH}} & =  -  \sum_{i,j, \sigma} t^{ij} c^\dagger_{i \sigma}c_{j\sigma}^{\phantom{\dagger}}+U \sum_{i} n_{i\uparrow}^{\phantom{\dagger}}n_{i\downarrow}^{\phantom{\dagger}},
\label{eq:hubbard}
\end{align}
where $c^\dagger_{i\sigma}$ and $c_{i\sigma}$ creates and annihilates a fermion of spin $\sigma \in \{\uparrow, \downarrow \}$ on lattice site $i$, respectively, and $n_{i\sigma} = c^\dagger_{i\sigma} c_{i\sigma}$ counts the number of fermions of the given spin on the corresponding site. The first term on the r.h.s. of Eq.~(\ref{eq:hubbard}) with prefactor $t^{ij}$ is the quadratic fermionic hopping term and the second term with prefactor $U$ is the quartic on-site interaction (also sometimes called coupling or density-density) term between fermions with opposite spins. Specifically, we will investigate two versions of the FHM: the model with only nearest-neighbor (NN) hopping terms, where $t^{ij} = t$ for all nearest neighbors $i$ and $j$ along the horizontal and vertical directions of the square lattice and $t^{ij}=0$ otherwise; and the model which additionally contains next-nearest-neighbor (NNN) hopping terms $t^{ij} = t^\prime$ along the two diagonals. 

\subsection{Encoding Restrictions}
\label{subsec:restrictions}

\begin{figure*}
    \centering
    \includegraphics[width=0.8\textwidth]{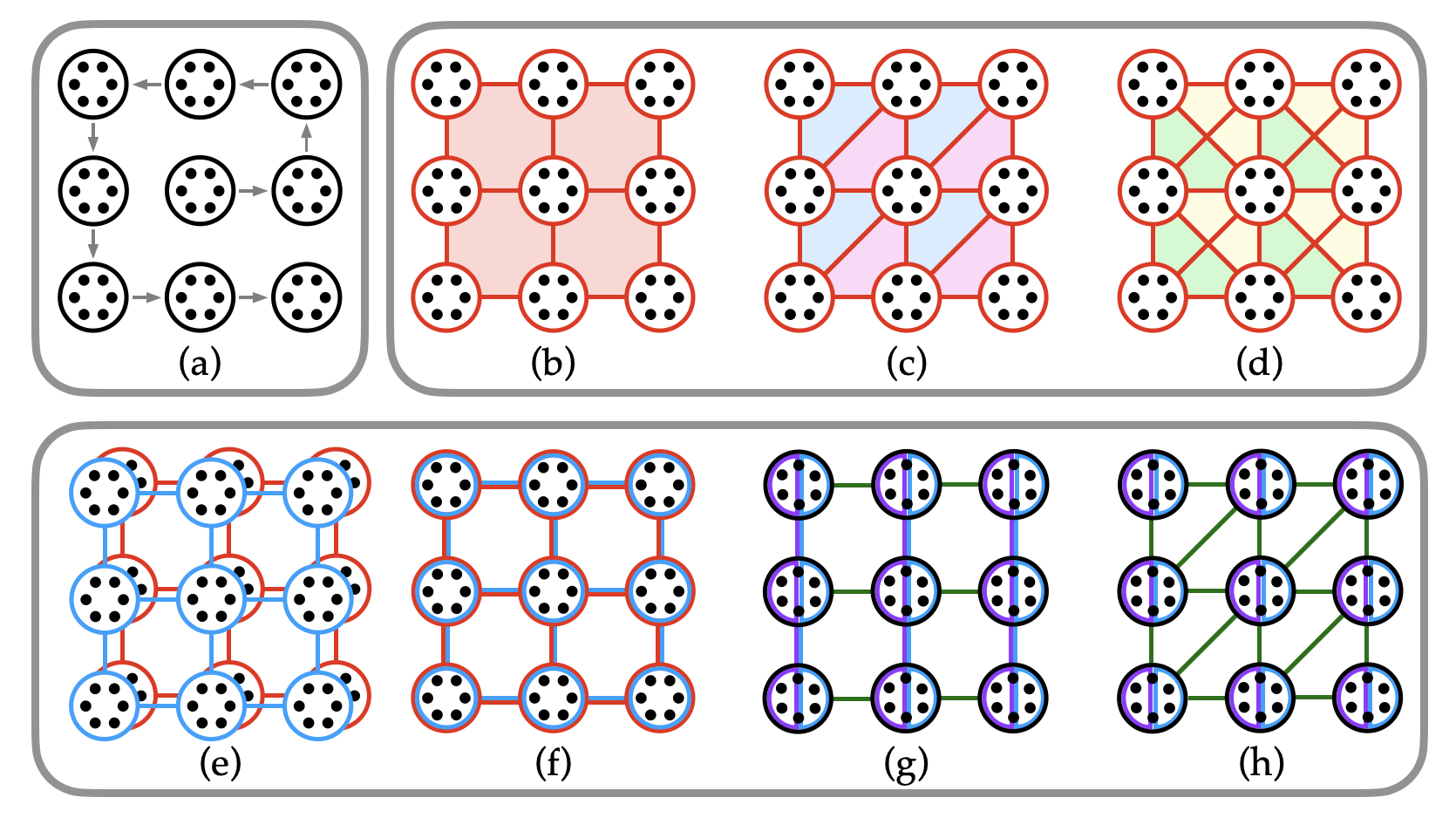}
    \caption{\justifying \textbf{Unit cells for fermionic encodings:} a) A 3-by-3 grid consisting of unit cells with up to 6 qubits per cell are efficiently encoded by two 64-bit integers. Arrows denote an (arbitrary) direction of the encoding b) If only horizontal and vertical edges are defined, then one unique stabilizer exists per unit cell (red square plaquettes). c) and d) If diagonal edges are additionally defined in one(both) directions then two(four) unique stabilizers exist as shown by triangular plaquettes in blue and pink (blue, pink, yellow and green). e) Two grids (red and blue) of the type of b), one per each spin. f) Both spins are embedded into the same grid. g) Doubling of the unit cell in the horizontal direction. h) Horizontal doubling of the unit cell with a vertical offset by one original unit cell.}
    \label{fig:unit_cells}
\end{figure*}

Due to translational and spin invariance of the FHM it is sufficient to define all operators for one unit cell consisting of a single fermionic mode of a given spin. This means that in the NN case we have to consider five distinct logical operators, four for the hopping operators acting on neighboring (horizontal or vertical) unit cells including their Hermitian conjugates and one additional interaction operator acting on two cells with fermionic modes of the same site and opposite spin. For the NNN case, there are four additional diagonal hopping terms, totalling 9 distinct logical operators.

Individual fermion-to-qubit encodings are derived using the formalism of edges (E) and vertices (V), as we have described in detail in Sec.~\ref{sec:Preliminaries}. It is straightforward to express the logical operators of the FHM in terms of $E$ and $V$. Specifically, for a pair of Hermitian conjugate hopping terms one has:
\begin{equation}\label{eq:hopping}
    c^\dagger_j c_k^{\phantom{\dagger}} + c^\dagger_k c_j^{\phantom{\dagger}} \rightarrow \frac{i}{2}(V_k-V_j)E_{jk},
\end{equation}
and for the on-site density-density term one has:
\begin{equation}\label{eq:on-site}
    n_j^{\phantom{\dagger}} n_k^{\phantom{\dagger}} \rightarrow \frac{1}{4} (1-V_j-V_k+V_j V_k).
    % (1-V_j)(1-V_k).
\end{equation}
From the definition in Eq.~\ref{eq:on-site} it is clear that the computational bottleneck will consist of implementing the last operator, $V_j V_k$, and we will thus disregard the remaining lower-weight operators in further analysis. 

We also do not necessarily have to define all edges directly, as they can be composed from other existing edges using the composite rule as long as an edge-path can be defined between the corresponding vertices, see Eq.~(\ref{eq:composite}).

We now have all the necessary ingredients to specify our approach to generating fermionic encodings, which are depicted graphically in Fig.~\ref{fig:unit_cells}. We study encodings generated from translationally invariant unit cells with up to six qubits per cell. Furthermore, in order to enforce a high degree of locality, we allow every operator to act on at most nine unit cells, arranged in a 3-by-3 square grid. This restriction on the qubit number is imposed in order to preserve computational performance, as we aim to efficiently encode all logical operators or stabilizers by two 64-bit integers in symplectic form, meaning one integer each for the $\bar{X}$ and $\bar{Z}$ parts of the operator. This means that all qubits from this grid can be encoded in a snake-like pattern, see Fig.~\ref{fig:unit_cells}a. 

Depending on the model of interest, one has to pick a set of edge and vertex operators, which are ideally close to the terms of the Hamiltonian to be simulated. The choice of this set will have strong influence the performance of fermion-to-qubit encoding, but the optimal set is generally not known a priori. Here, we consider three different edge connectivity graphs: NN square (horizontal and vertical edge operators), triangular (one additional diagonal edge operator) and NNN square (both diagonal edge operators) (Fig.~\ref{fig:unit_cells}b-d). If one intents to simulate a NNN square Fermi-Hubbard model using encodings with lower edge connectivity, all missing edges are constructed using the composite rule of Eq.~(\ref{eq:composite}). As we know from Eq.~(\ref{eq:stabilizer}), every closed loop of edge operators corresponds to a stabilizer operator in a given encoding graph. This means that our choice of encodings results in one square stabilizer per unit cell for the NN square encodings (red plaquettes in Fig.~\ref{fig:unit_cells}b), two triangular stabilizers for the triangular encodings (light blue and pink triangles (Fig.~\ref{fig:unit_cells}c) and four triangular stabilizers for the NNN square encodings (the stabilizers of Fig.~\ref{fig:unit_cells}c as well as the yellow and green triangles of Fig.~\ref{fig:unit_cells}d). Ideally, we would like the resulting stabilizers to be as low-weight as possible. Our choice of encoding graphs, especially for the triangular and NNN square cases, is thus deliberate as they only act on three vertices. Additionally, the increased number of stabilizers might provide additional \emph{coverage} of qubits and potentially improve the distance of our encodings. In terms of qubit connectivity we, for the time being, allow any qubits within our 3-by-3 grid to interact and will only study and restrict the connectivity of encodings at a later point, in Section~\ref{sec:Codesign}. 

Since we are interested in simulating the FHM with two spin-modes per lattice site, we need to take into account the density-density operators defined between them. The easiest encoding strategy is to simply define two copies of a given encoding for only one spin-type (Fig.~\ref{fig:unit_cells}e). This works, because the coupling term is given exclusively in terms of vertex operators, meaning that no edge needs to be defined between fermionic modes of different spin-types. We want to note that, contrary to some previous studies, we do not consider the $t$-$J$ Hubbard model here, for which interaction terms are defined between NN modes connected by an edge operator. 

As an alternative to having two copies of a one-spin encoding, we also search for mixed encodings where the vertex operators for modes of both spin-types associated with a lattice site are acting on a common set of qubits within a single unit cell (Fig.~\ref{fig:unit_cells}f). In similar fashion, we allow for two horizontally neighboring vertices of the same spin to be acting on qubits of the same unit cell (Fig.~\ref{fig:unit_cells}g and Fig.~\ref{fig:unit_cells}h). This effectively renders the square lattice of edges and vertices bipartite in the horizontal direction, thus allowing for more varied encodings. The difference between Fig.~\ref{fig:unit_cells}g and Fig.~\ref{fig:unit_cells}h is that in the latter we introduce an additional vertical offset between doubled unit cells. As a consequence, the lattice is also effectively bipartite in the vertical direction. We note that whilst such encodings allow to have more degrees of freedom compared to the single spin mode per unit cell ones, they also come with a significant computational overhead, as they contain double the number of distinct logical operators and enlarged unit cells. We will give more specific examples of existing encodings from literature, which correspond to such cases in the following sub-section. 

\subsection{Local Fermion-to-qubit encodings from literature}
\label{sec:literature_mappings}

Whilst a large number of different local fermion-to-qubit encodings with differing advantageous characteristics have been proposed to date, it has been shown that any two local encodings can be transformed into each other using Clifford transformations \cite{chen2023equivalence}. Nevertheless, it is in many cases necessary to break translational symmetry in order to successfully transform between two different encodings. 

Some classes of local encodings have been designed with the purpose of maximally reducing either the logical operator weights, the number of two-qubit gates of the circuit depth \cite{derby2021compact, verstraete2005mapping, algaba2023low} and, as a consequence, the distance of most of these encodings is only $d=1$. A number of recent publications have taken the alternative route of trying to find encodings with higher distances $d>1$ instead. In Ref.~\cite{chien2022optimizing} the authors identified multiple $d=2$ encodings with low operator weights and tailored these to particular quantum hardware connectivity graphs. In Refs.~\cite{jiang2019majorana, setia2019superfast, chen2022error, landahl2023logical} the authors proposed $d=3$ encodings and Ref.~\cite{chen2022error} went to much higher distances of $d \leq 7$ for encodings derived from exact bosonisation~\cite{chen2018exact}. We will use the aforementioned encodings as benchmarks against which we will compare the performance of our novel encodings. 

All of the aforementioned encodings can be in theory \emph{rediscovered} using the methods we will present in detail in the next section, as long as they fit within the restrictions described in Section~\ref{subsec:restrictions}. For example, we found nearly all of the $d\leq3$ encodings, e.g. those from  Refs.~\cite{verstraete2005mapping,chen2018exact,derby2021compact,chien2022optimizing,algaba2023low}. Notable exceptions are the encoding from Ref.~\cite{jiang2019majorana}, for which one would need a construction consisting of four distinct unit cells, and the encoding from Ref.~\cite{landahl2023logical}, where the approach is to build an encoding from introducing \emph{defects} in the surface QEC code which results in way too large unit cells. 

\subsection{Brute-Force Search}
\label{sec:bruteforce_search}

The first method for identifying new fermion-to-qubit encodings is a brute-force search algorithm, which iterates through all possible Pauli string definitions of logical operators whilst verifying that commutation relations between operators are correct for all operators within and between translationally invariant unit cells. The most efficient way to do so is by comparing the matrices $\sigma$ and $\tau$ with Laurent polynomial entries, as described in Section~\ref{sec:Preliminaries} and similar to the  approach used in Ref.~\cite{chien2022optimizing}. 

We choose to fix the order in which we search for logical operators to: vertex, horizontal edge, vertical edge, diagonal edge, second diagonal edge. In encodings with two fermionic modes per unit cell we first search for operators of one and then the other mode. Whenever a valid operator is found, we proceed to the next one. If the search for an operator finishes by iterating through all allowed Pauli strings or no valid operator is found, the algorithm returns to the previous logical operator and then continues the search there. 

In order to improve efficiency by reducing the number of possible Pauli string combinations for logical operators, we have added a few additional restrictions to the search space. First, we enforce that the vertex must act on at least one qubit of the central unit cell of the 3-by-3 cell lattice. Similarly, every edge must act on at least one qubit in the central unit cell as well as at least one qubit in the unit cell containing the other vertex which it anticommutes with. Further, we introduce an (arbitrary) order in which qubits of unit cells are acted upon. A qubit can only be acted upon if every previous qubit is being acted on in at least one unit cell by at least one logical operator (which generalizes to all other unit cells by translation). Similarly, a new qubit is always first acted upon with a Pauli $Z$ operator and only then, successively with a Pauli $X$ and Pauli $Y$ for further logical operators. This way we limit the number of similar encodings that would be generated due to the exchange symmetries between Pauli operators.

Since we are only interested in well-performing encodings, we can also limit the maximum weights of stabilizers and logical operators, the minimum distance and in that case also the minimum weights of logical operators. Given the fact that one, in practice, never implements edge operators on a quantum computer, but rather hopping operators, which occur in fermionic Hamiltonians, we can limit the minimum/maximum weights of hoppings instead of edges. Here, we make the distinction between limiting the weight for either just NN or both NN and NNN hoppings. Vertices, in contrast, do occur in the Hamiltonian as they correspond to density operators. 

The higher the maximum allowed weight of logical operators is set, the higher the number of possible Pauli strings and, consequently, the number of different encodings. In practice, the number of options can be too large for the brute-force algorithm to go through all of them. In this case, it is useful to use a version of the algorithm which only accepts each new valid logical operator with a given (arbitrarily chosen) probability. We find that within a given set of restrictions, similar encodings tend to appear relatively often (despite our best efforts to limit the symmetries in the generation thereof). Therefore, the search algorithm is successful at finding most encodings of interest even with low acceptance probabilities of $1-10\%$.

For all new encodings we compute the distance, the maximum stabilizer weight and the average weight of logical operators including either NN or both NN and NNN hopping operators. The distance is computed by generating all possible Pauli string operators up to a certain weight, which act on the 3-by-3 unit-cell grid and checking whether they either are a stabilizer or anticommute with at least one stabilizer. If none of the two statements hold, this means that the operator is a logical operator and the code distance is limited by its weight. If all operators up to a given weight pass this test, the minimum distance is at least equal to this weight. Similarly to Ref.~\cite{chen2022error}, our algorithm which computes the distance of an encoding scales exponentially with the allowed operator weight as well as the maximum number of qubits.

We then instruct the search algorithm to output all new encodings as long as they match or improve upon all previously found encodings in at least one of these four metrics. We want to stress that, at this point, no connectivity considerations are taken into account. 

\begin{figure*}
\centering
\includegraphics[width=1\textwidth]{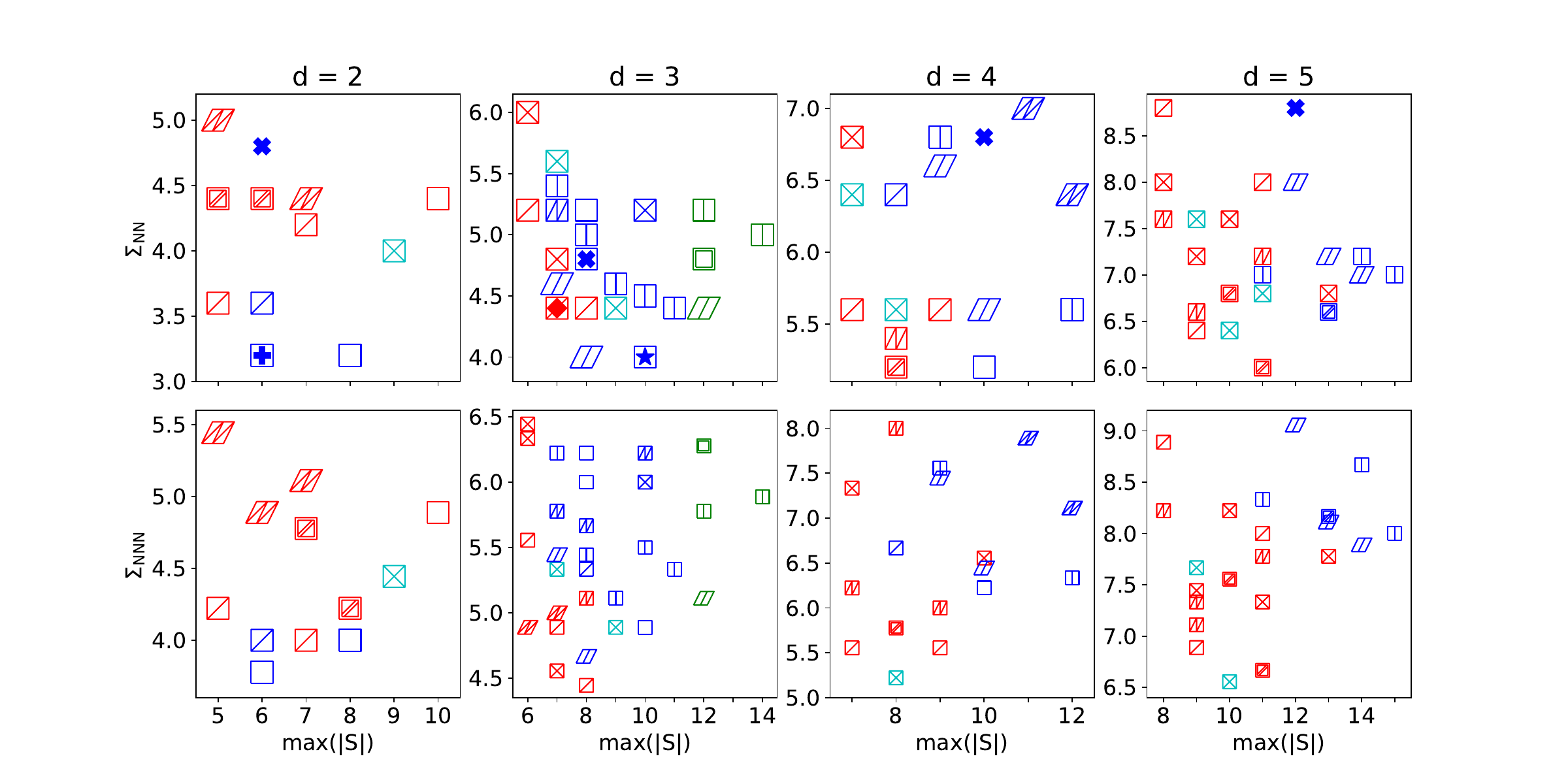}
\includegraphics[width=0.8\textwidth]{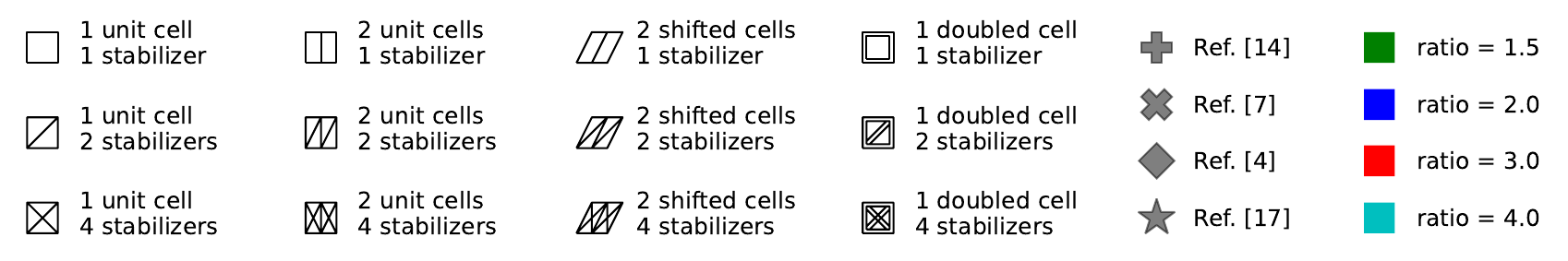}
\caption{\justifying \textbf{Fermion-to-qubit encodings} with distances $d \in \{ 2,3,4,5 \}$ as functions of their average operator weight, for operators from the Fermi-Hubbard model with nearest- ($\Sigma_{\text{NN}}$) and next-nearest- ($\Sigma_{\text{NNN}}$) neighbor hopping terms, and as a function of the maximum stabilizer weights ($\operatorname{max}(|S|)$). Each data point is an individual encoding corresponding to a construction given by the legend. For comparison, we show existing encodings from literature. The qubit-to-fermion ratio is indicated in color.}
\label{fig:combined_plot_2345}
\end{figure*}

\subsection{Clifford Search}
\label{sec:clifford_search}

We used the brute-force method described in the previous section to find a number of \emph{base} fermion-to-qubit encodings. Now, we can use these base encodings as input, which is used to derive new encodings through Clifford deformations. Since the elements in the Clifford group are unitary transformations and commutators are invariant under these, the commutation relations of a given base encoding are preserved after the application of Clifford gates. For every distinct pair of qubits, there are $256$ possible Pauli operators that can be reached by applying Clifford gates to them. Given the large number of possible qubits pairs, however, we need to develop a technique that produces interesting new encodings by taking into consideration the unit cell connectivity whilst using only a reduced fraction of all possible Clifford gates.

In principle, one could also consider the use of generalized local unitary (GLU) operators~\cite{Chen_2023} for this purpose. Since adding or removing an extra ancilla qubit that is disentangled from the rest of the system is an operation that does not affect the underlying state, then GLU operators can be applied to the base encoding to produce new encodings with a different number of qubits. As was shown in Ref.~\cite{Chen_2023}, GLU operators can be used to generate new encoding as well as to show equivalence between existing encodings. However, we are interested in encodings that do not arbitrarily increase the fermion-to-qubit ratio, and therefore this strategy is not pursued here.

Let us now focus on our methodology for reducing the number of used Clifford gates. Since the initial and final encodings are translation invariant, let us consider the first unit cell only, but our method is also applicable to other unit cells, as we are going to describe later on. The methodology is two-fold. Firstly, we choose which single-qubit Clifford gate is going to be applied onto each qubit of the first unit cell of the base encoding. Since there are $2^{3n}$ single-qubit Clifford gates for a $n$-qubit system but no reason to avoid using any of them, we randomly choose a certain amount of single-qubit Clifford gates to apply. This process is repeated for every qubit in the first unit cell and the chosen single-qubit Clifford gates are added to the set of possible Clifford gates. Note that single-qubits Clifford gates can not increase the weight of logical operators and therefore the distance. They can, however, reduce these weight, which ultimately is necessary for increasing \emph{mixing} of the algorithm, which in turn allowed us to identify better encodings.  

Secondly, we consider CNOT gates between qubits in the first unit cell and any other unit cell which are connected through a common edge operator and add these to the set of possible Clifford gates. In the case of qubits within a single unit cell, we simply include all possible CNOT gates. For two qubits from different unit cells, we take into consideration the type of edges in the fermion-to-qubit encodings. Suppose two unit cells $i$ and $j$, with $n_c$ qubits per unit cell, are connected. We label their qubits as $q_{i,l}$ and $q_{j,k}$, respectively, for $l,k=1,\ldots,n_c$. Randomly, we add a predetermined number of pairs $\{q_{i,p},q_{j,p}\}$, for $p\in\{1,\ldots,n_c\}$, to the set of possible Clifford gates. After have created this set, we run all combinations of elements in it. Since the size of possible Clifford gates is significantly smaller then the size of the Clifford group, the algorithm can be executed in a reasonable time.

To preserve the translational invariance of the base encoding, when a sequence of Clifford gates is applied within the first unit cell or between the first unit cell and any unit cell connected to the first cell by an edge operator, we also apply the same Clifford gate to every other unit cell that is translationally equivalent to it. This operation guarantees that the new encoding is also translationally invariant.

\section{Results}
\label{sec:Results}
\begin{figure}
% \centering
\includegraphics[width=0.8\columnwidth]{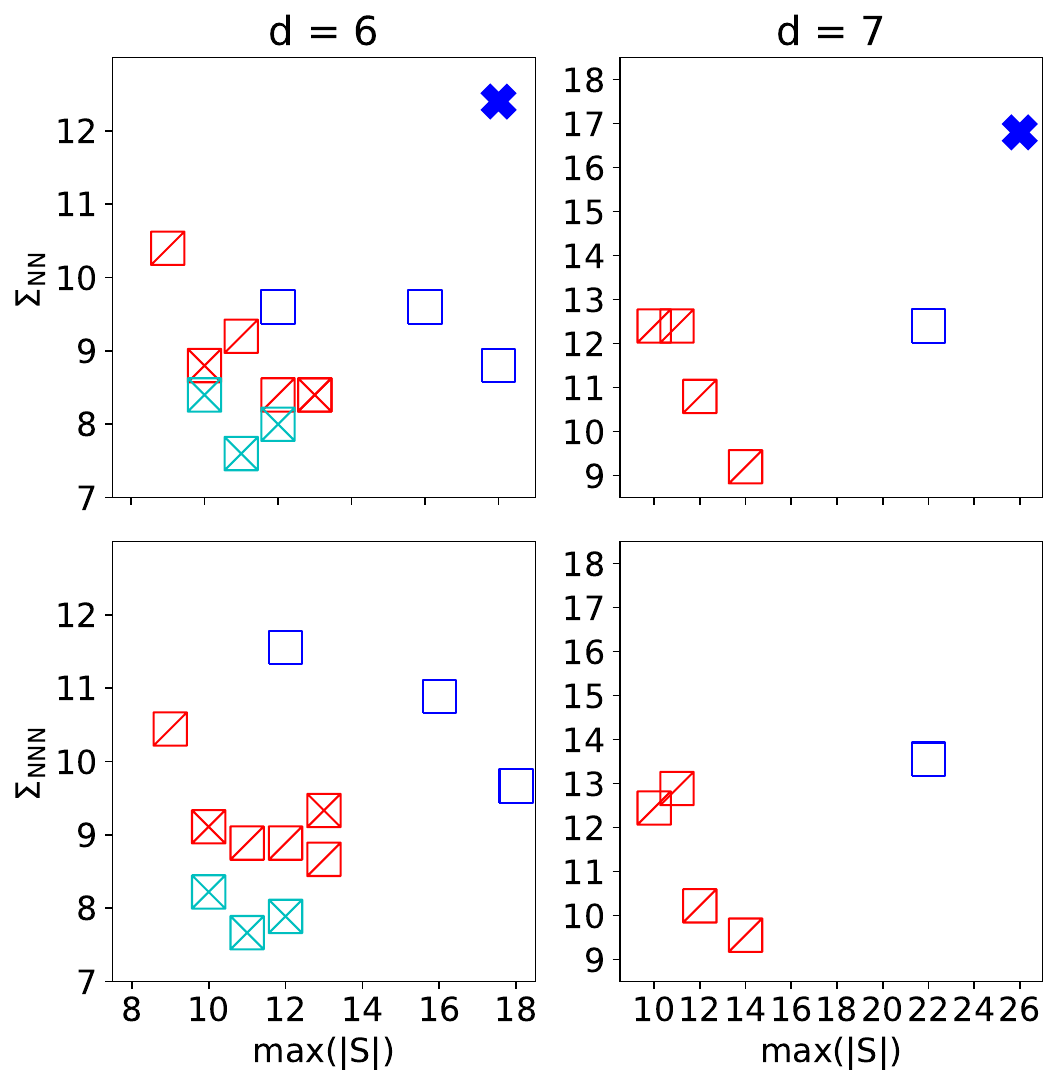}
\caption{\justifying Fermion-to-qubit encodings with distances $d \in \{ 6,7 \}$ as functions of their average operator weigh, for operators from the Fermi-Hubbard model with nearest- ($\Sigma_{\text{NN}}$) and next-nearest- ($\Sigma_{\text{NNN}}$) neighbor hopping terms, and as a function of the maximum stabilizer weights ($\operatorname{max}(|S|)$). Each data point is an individual encoding corresponding to a construction given by the legend of Fig.~\ref{fig:combined_plot_2345}. For comparison, we show existing encodings from literature. The qubit-to-fermion ratio is indicated in color.}
\label{fig:combined_plot_67}
\end{figure}

In this section we will present an overview of the encodings we were able to identify by combining the two algorithms from Sections \ref{sec:bruteforce_search} and \ref{sec:clifford_search}. Our calculations have been performed on a single CPU and the total run-time was of the order of 20 hours. 

The highest code distance encodings we found using the brute-force search were $d=4$. The computational bottleneck in this algorithm came mainly from operations on Laurent polynomial matrices as well as computations of the distance. We note that our computational search space was significantly larger than what was reported in Ref.~\cite{chien2022optimizing}, where the authors only considered distance $d=2$ encodings with a maximum weight of 3 for logical edge and vertex operators. Another difference is that we focused here on the weight of hopping operators rather than edge operators, and took into account stabilizer weights in our evaluation of encodings. 

In a second stage, we took the best-performing encodings from the brute-force search and used them as starting-point encodings for the Clifford search algorithm. Even though all of these encodings can be transformed into each other using Clifford transformations, the number of steps between them can be prohibitive. Within the Clifford search algorithm, we first scanned through all the encodings generated by up to three Clifford deformations. Then we generated of the order of $10^7$ random Clifford sequences with a varying number of up to 15 deformation steps. The maximum code distance found for encodings was $d=7$. For the Clifford search algorithm the bottleneck was primarily in the computation of the distance, as investigating a single encoding of distance $d=7$ which acts on unit cells containing 4 qubits can take up to about one minute. For this reason, we limited ourselves to this maximum distance for 4-qubit unit cells and only went up to distance $d=5$ for unit cells containing 6 qubits (in most cases corresponding to encodings with two fermionic modes per unit cell). One could theoretically search for encodings of even higher distances and qubit-to-mode ratios by running the Clifford search algorithm in parallel on high-performance computing clusters.

A birds-eye view of (all the reasonably interesting) encodings we have identified in this work is presented in Fig.~\ref{fig:combined_plot_2345} for encodings with distances $2\leq d \leq5$ and Fig.~\ref{fig:combined_plot_67} distances $6\leq d \leq7$. We plot all encodings in terms of their maximum stabilizer weight, $\max{(|S|)}$, and their average logical operator weight, averaged over vertices, NN hoppings, and excluding/including (top/bottom) NNN hoppings ($\Sigma_{\text{NN}}$/$\Sigma_{\text{NNN}}$). Further, we indicate the qubit-to-mode ratio with colors and the type of encoding (see Fig.~\ref{fig:unit_cells}) using different symbols. For comparison, we have also added encodings from Refs.~\cite{jiang2019majorana, setia2019superfast, chien2022optimizing, chen2022error} in the top (NN only) plots, where such results exist.

For distance $d=1$, the optimal encoding in terms of the average weight $\Sigma_{\text{NN}}$ and $\Sigma_{\text{NNN}}$ have been identified in Ref.~\cite{derby2021compact} and Ref.\cite{verstraete2005mapping}, respectively. Ref.~\cite{derby2021compact}, in particular, has the added advantage of having a fermion-to-qubit ratio of only $r=1.5$.  However, we found that ratio $r=1.5$ encodings are performing worse than the higher ratio alternatives (we have specifically added a number of $r=1.5$ encodings into the $d=3$ plots) for higher distances. In general, we see that, as the distance increases, increasingly higher ratio encodings are becoming preferable, which intuitively is logical, as there are more qubits available to resolve commutation relations. The same also seems to be true for encodings with increased connectivity when NNN hoppings are included into the logical operators metric ($\Sigma_{\text{NNN}}$). One immediate question is why higher ratio encodings do not perform at least as well as low ratio ones, especially given their additional degrees of freedom. Since we only consider encodings which act on all available qubits, however, this restriction leads to the lower performance observed in these higher ratio encodings.

For $d=2$, our best encoding corresponds to the one identified in Ref.~\cite{chien2022optimizing}, which is not surprising as our brute-force algorithm is roughly equivalent to theirs and the search space is small enough to exhaustively investigate all possible encodings. The encoding has a ratio of $r=2$, a maximum stabilizer weight of 6, and average logical operator weight of close to 3. The encoding is optimal in the sense of both $\Sigma_{\text{NN}}$ and $\Sigma_{\text{NNN}}$. Here, we also show the original exact bosonisation (EB) encoding from Refs.~\cite{chen2018exact, chen2022error}, which however does not perform very well in comparison to the optimal one. In terms of purely looking at the stabilizer metric, we find an $r=3$ encoding with one diagonal edge, and $\max{(|S|)}=5$.

We now turn to $d=3$ encodings. Such encodings have been previously reported in Refs.~\cite{jiang2019majorana, setia2019superfast, chen2022error}. We find that the EB encoding, from Ref.~\cite{chen2022error}, is outperformed by the generalized superfast encoding (GSE \cite{setia2019superfast}) in terms of both logical operator and stabilizer weights. We want to note that Ref.~\cite{setia2019superfast} only provided the definitions in terms of Majorana operators, but not edge and vertex operators, the choice of which will influence the weights of stabilizers. For this comparison, we have thus chosen the best edge combination corresponding to the Majorana operators provided by Ref.~\cite{setia2019superfast}, as found by our algorithms. The Majorana loop stabilizer code (MLSC \cite{jiang2019majorana}) performs even better in terms of $\Sigma_{\text{NN}}$, but has stabilizers of higher maximum weight of 10 as compared to 7 for GSE. In fact, we found an encoding with similar properties, which however, only requires to define one distinct unit cell instead of four for MLSC. We also found an encoding which is an improvement upon both, as it has the same value for $\Sigma_{\text{NN}}$, but $\max{(|S|)}=8$. This is a shifted doubled unit cell encoding, see Fig.~\ref{fig:unit_cells}h. In the bottom (NNN) plot, we observe that encodings with $r=3$, and one or two diagonal edges are scoring better than the aforementioned $r=2$ encoding.  

For distances $4\leq d \leq7$, the only previously reported results are generalizations of the EB encoding from Ref.~\cite{chen2022error}. We see from Figs.~\ref{fig:combined_plot_2345} and \ref{fig:combined_plot_67} that these encodings are seriously outperformed by the ones found in this study in both the logical operator and stabilizer metrics. For these distances, the best performing encodings we found have ratios of either $r=3$ or $r=4$ and have high connectivity, containing either one or two diagonal edges. Additionally, we find for $d=4$ and $d=5$ well-performing encodings with doubled unit cells, meaning two modes of opposite spins defined within a single unit cell. This is mainly due to the fact that we consider the density-density on-site interaction operator from the Fermi-Hubbard model, rather than its t-J model counterpart, where the interaction is defined between neighboring sites. For $r=4$, we observe that a family of encodings with both diagonals performs well, especially for $\Sigma_{\text{NNN}}$ and $d=6$, where these encodings outperform all $r=3$ ones. Because of the high computational cost, we only searched for encodings with up to $r=3$ for $d=7$. It is rather likely that one would be able to find good $r=4$ encodings here as well. In absence of these, we find $r=3$ encodings with one diagonal edge to be the best. 

\begin{figure}
\centering
\includegraphics[width=0.8\columnwidth]{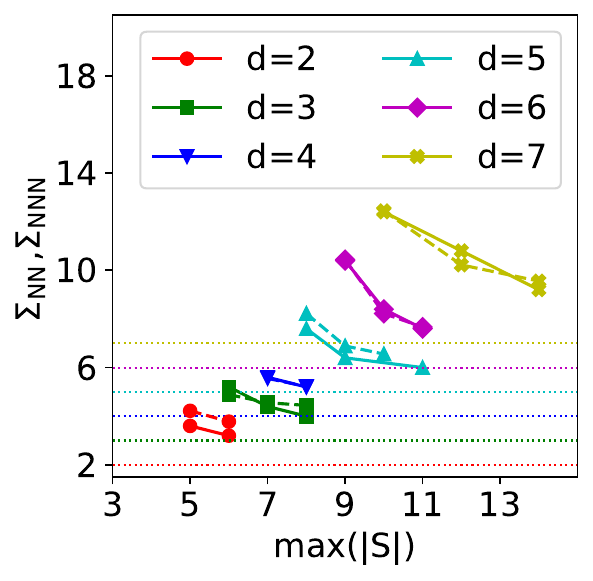}
\caption{\justifying Optimal fermion-to-qubit encodings for distances $2\leq d \leq 7$ in terms of the combination of small maximum stabilizer weight and small average operator weights. Full and dashed lines correspond to the operators from the Fermi-Hubbard model with NN and NNN hopping terms, respectively. Dotted lines are theoretical lower bounds of operator weights for given distances.}
\label{fig:best_encodings}
\end{figure}

In Fig.~\ref{fig:best_encodings} we summarily plot the best encodings of all distances. Within a given distance, an encoding was added to this plot only if it either has a lower $\Sigma_{\text{NN(N)}}$ than all encodings which have lower $\max{(|S|)}$ than itself, or vice-versa. The near-equal spacing (with the exception of the $d=7$ curve) suggests that the encodings we found are close to optimal, at least within the restrictions introduced on the search algorithms. We also see that the difference between the average logical weights of $\Sigma_{\text{NN}}$ and $\Sigma_{\text{NNN}}$ is nearly negligible, which is promising for the simulation the Fermi-Hubbard model with further-neighbor hopping terms. Such models have a strong connection to cuprate physics and are harder to simulate classically than the plain Fermi-Hubbard model with only NN hopping terms.In Fig.~\ref{fig:best_encodings}, the minimum average logical operator weights $\Sigma_{\text{NN(N)}}$, as prescribed by the distance, are given by dashed horizontal lines. Whilst there is a minimum for the logical operators, the stabilizer weights can, from a quantum error-correcting point-of-view, be arbitrarily small.
What we observe, however, is a trade-off between the weights of logical operators and the weights of stabilizers, both of which slowly grow with increasing distance. This poses a significant challenge for the measurement of stabilizers of higher-distance encodings on quantum processors with limited qubit connectivity, but can be mitigated by using multiple syndrome qubits and/or SWAP gate operations. Interestingly, we always observe at least difference of three between the distance of an encoding and the maximal weight of stabilizers, but we do not have a clear explanation for this finding.

The details of a hand-picked selection of the encodings from Figs.~\ref{fig:combined_plot_2345} and \ref{fig:combined_plot_67} (including qubit connectivity structures discussed in the next section) 
are presented in appendix \ref{appendix}.

\section{Co-design Hardware Implementations}
\label{sec:Codesign}
\begin{figure*}
    \centering
    \includegraphics[width=0.25\textwidth]{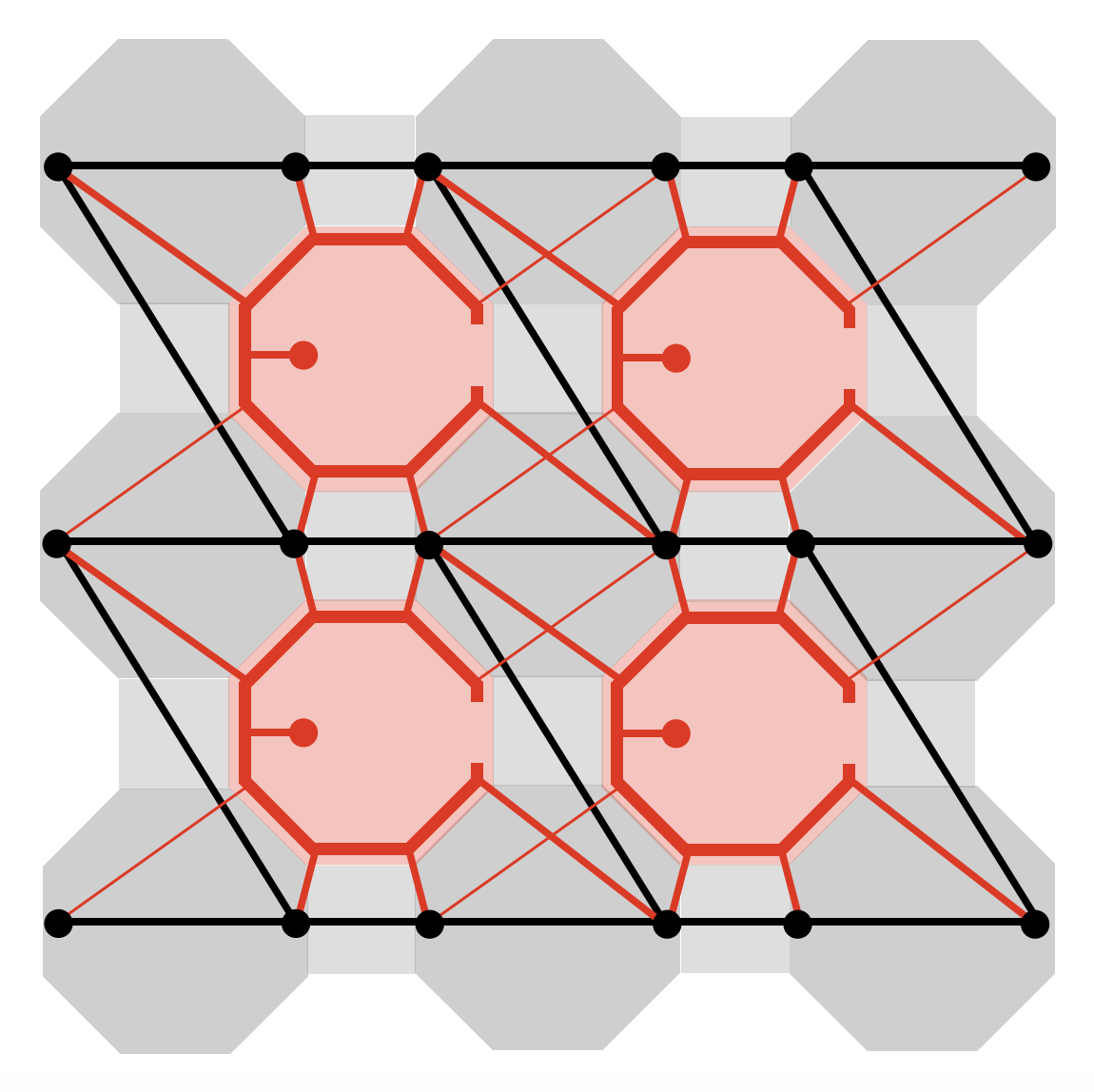}
    \includegraphics[width=0.25\textwidth]{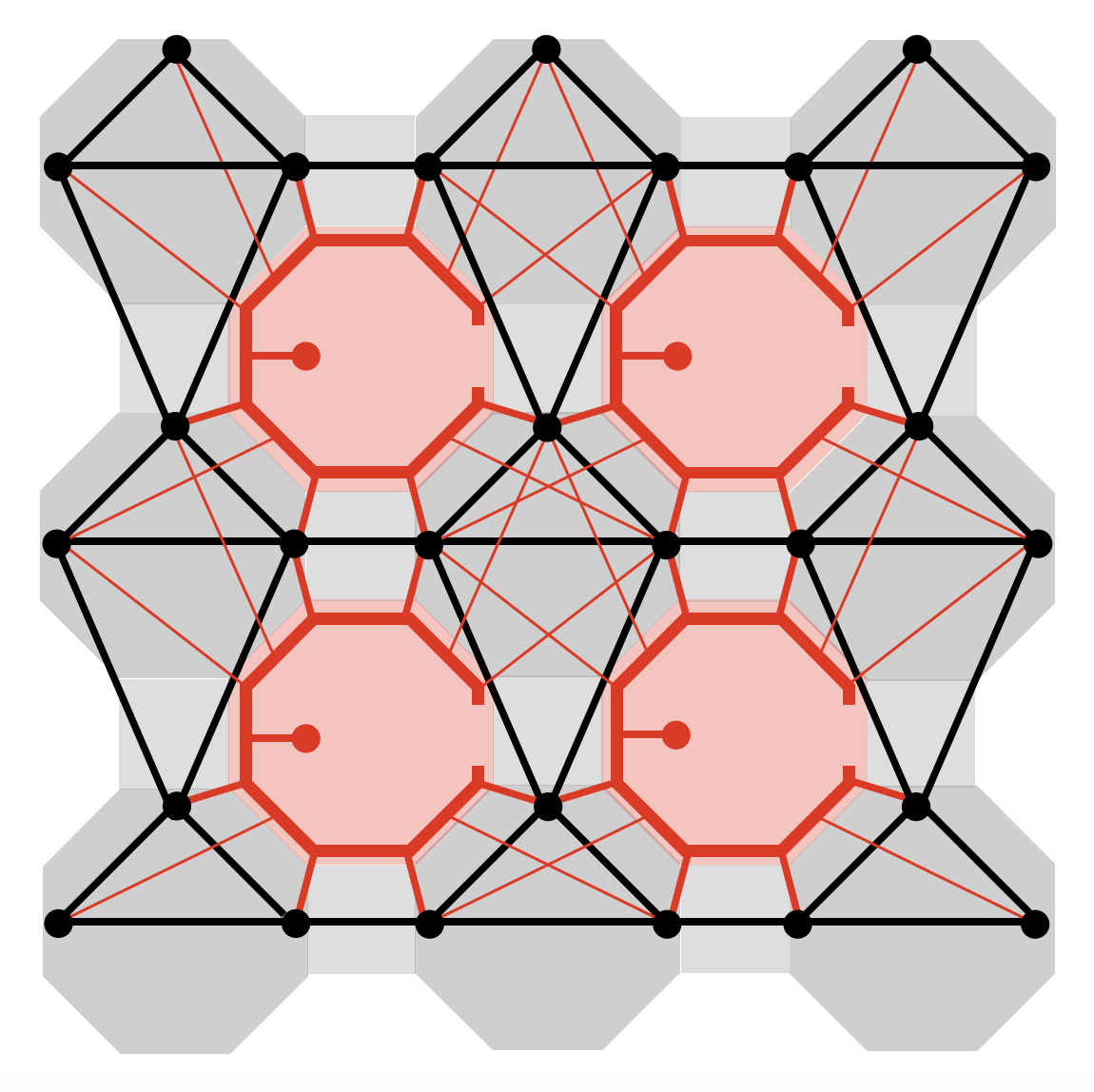}
    \includegraphics[width=0.25\textwidth]{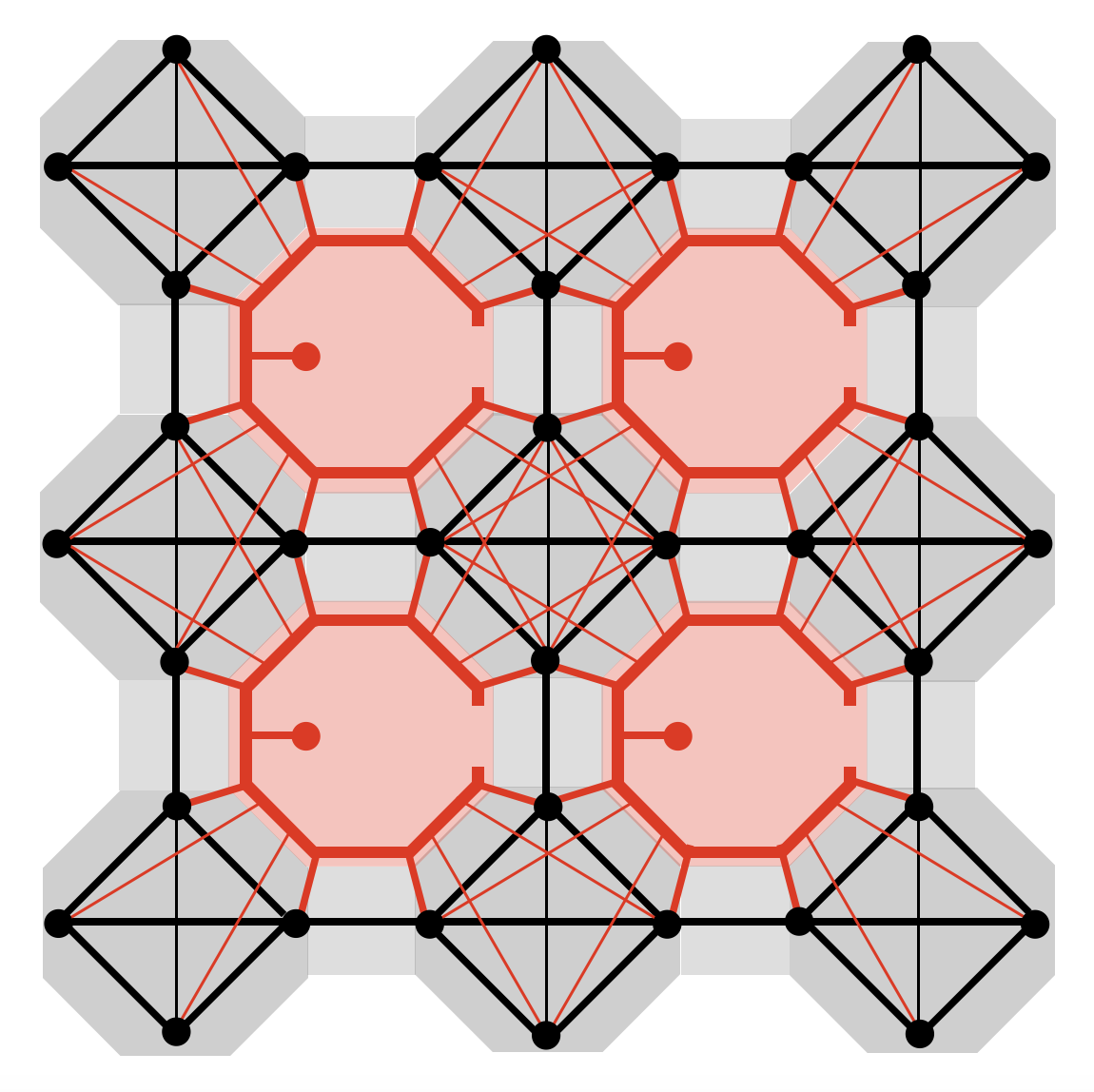}
    \caption{\justifying Sketches of co-design chips allowing the implementation of some encodings with fermion-to-qubit ratios of 2,3 and 4 (left to right). Data/syndrome qubits are represented by black/red dots. Black lines are tunable couplers between data qubits. Red lines connect data qubits to central resonators. Thick/thin lines indicate the positioning of couplers on the first/second layer of the flip-chip.}
    \label{fig:codesign}
\end{figure*}

In this section, we will focus on the realistic quantum hardware implementation of the fermion-to-qubit encodings found in this work. For quantum devices with all-to-all connectivity one does not need to take the qubit connectivity graph into consideration. Many quantum technologies, however, only provide a limited number of connections per qubit and routing information around using SWAP gates introduces additional errors thus deteriorating overall performance of algorithms. In the latter case, it is crucial to optimise the qubit connectivity (hardware) graph.

One can use our encodings to not only correct errors using mid-circuit measurements in the traditional QEC sense, but also for NISQ devices that may come with a prohibitive syndrome-measurement cycle overhead. As an alternative to QEC, one can implement various error mitigation schemes to improve the performance of algorithms in NISQ devices, albeit at an exponentially scaling overhead in the number of circuit runs. Nevertheless, such a quantum error mitigation approach might be viable for moderately sized quantum computers, short circuits and good gate fidelities. For example, one can do a ``one-shot’’ QEC cycle, where stabilisers are measured at the end of a circuit and the results are post-selected. This has the up-side that whilst one can only correct errors up to $(d-1)/2$, it is possible to detect errors up to $d-1$. Another option is to project the results onto stabiliser subspaces, as was proposed in Ref.~\cite{mcclean2020decoding}.  Notably, contrary to performing QEC, such approaches do not require additional syndrome qubits to be available on the quantum hardware.

Given a particular encoding, our primary aim will thus be to identify \emph{optimal} connectivity structures that allow for all logical operators from a given Hamiltonian to be implemented without the addition of SWAP gates, in our case that of the Fermi-Hubbard model (see Eq.~\ref{eq:hubbard}). Here, we will make the distinction between investigating the \emph{pure} Fermi-Hubbard model with only NN hopping terms, and the model with both NN and NNN terms present in the Hamiltonian. The Fermi-Hubbard model on most other two-dimensional lattices of interest can then be constructed by including a subset of the NN and NNN hopping terms \cite{algaba2023low}. Furthermore, we will define optimal connectivity structures as those that have a planar qubit connectivity graph and the lowest possible maximal number of connections for any given qubit. These restrictions are inspired by e.g. current superconducting quantum chips, where the number of couplers per qubit is limited to 3-4, but likely extendable to about 8 without suffering from a severe loss in performance. Similarly, non-planar connectivity graphs require the use of either air-bridges or the introduction of couplers between different sides of flip-chip architectures as was shown in Ref.~\cite{field2023modular}. 

We thus aim to implement an algorithm which considers all couplings for pairs of qubits either belonging to the same unit cell, or to two neighboring unit cells in the horizontal, vertical or diagonal directions and no couplings between further-apart unit cells are allowed. in order to avoid confusion with the vertex and edge operators we have previously introduced, we will denote couplings as links and qubits as nodes in the hardware graph. Our encodings are translationally invariant, which means that for any link added to the graph, all translated links must likewise be added to it. Whilst this restriction reduces the number of possibilities significantly, we still have $n_{\text{links}} = 9/2 \, n_c^2-n_c/2$ different links to choose from for a unit cell of $n_c$ qubits. Since the number of possible hardware graphs scales as $2^{n_{\text{links}}}$, an exhaustive search becomes quickly computationally prohibitive, even for a unit cell with $n_c=3$ where $n_{\text{links}}=39$. This number can be somewhat further reduced by only considering links between pairs of nodes jointly belonging to at least one logical operator, but, in practice, we found that this only yields a small computational improvement. We therefore resort to a heuristic approach which iteratively scans through remaining available links and picks one which connects most previously disconnected components of individual logical operators. Note that the set of all translated links is added at the same time, which can affect multiple logical operators within different unit cells. Additionally, we set a penalty for links which increase the maximum connectivity of any given qubit. For each encoding we first try to identify a planar connectivity graph and only if no solution is found we progress to searching for optimal non-planar graphs. 

In Tables \ref{table:enc1}, \ref{table:enc2}, \ref{table:enc3} and \ref{table:enc4} of Appendix \ref{appendix}, we provide results for a hand-picked set of encodings with favourable properties and will only provide general statements about our findings here. In general, we observe that maintaining planarity as well as low connectivity becomes increasingly hard for higher distance codes. Further, we identify multiple encodings where we can maintain planarity and low connectivity for the NN graph, but not for the NNN graph. Nevertheless, we managed to identify planar NN graphs with a maximum connectivity of 6 even for $d=7$ encodings, whilst in the case of NNN this was only possible for up to $d=4$ encodings. For lower distance encodings, $d\leq4$ we found that maintaining planarity as well as reasonable connectivity, meaning less or equal than four connections per qubit, is in fact possible for many encodings.  One particularly promising $d=4$ encoding with four qubits per unit cell that we found is in fact planar and has only connectivity 3 for the NN graph, whilst keeping all logical operator weights below 6 and all stabilizer weights below 8. 

Keeping the data qubit connectivity graph planar also enables for the syndrome qubits to be coupled to data qubits whilst maintaining a bi-planar graph for both types of couplers combined. In Appendix \ref{appendix}, we represent any number of syndrome qubits used for the measurement of stabilizers by single red dots and the necessary connectivity for measuring stabilizers by red links. We note that one could, in principle, also use one syndrome qubit per stabilizer and even use additional flag qubits to capture hook errors which occur during a measurement cycle. All stabilizers are local and, ideally, we want them to only act on found neighboring unit cells. Since in the worst case, a syndrome qubit has to couple to all qubits of its surrounding unit cells on a square unit cell lattice, it is possible to resolve these connectivities without breaking bi-planarity as long as the number of qubits per unit cell does not exceed four. This is done by coupling to two qubits on the same plane as the data qubit couplers without crossing them and then using a second plane to couple to the remaining two qubits in a spiral arrangement. In the same way, one would be able to resolve the coupling for up to six qubits per unit cell on a honeycomb unit cell lattice, and up to three qubits per unit cell for a triangular lattice. 

Other than planarity, the main obstacle to physically realising these encodings in real quantum devices is the necessary connectivity of syndrome qubits. That can be as high as 12 for any given single stabilizer of higher-distance encodings and upper-bounded by $4n_c+1$ if we aim to measure multiple stabilizers of the types c-h in Fig.~\ref{fig:unit_cells} with the same syndrome qubit. This connectivity is beyond the capability of commonly implemented superconducting qubits such as transmons or fluxoniums. However, this can be achieved using resonator elements, which can be connected to up to around 20 tunable couplers \cite{song2019generation}. That is sufficient for all the encodings we identified in this work since for unit cells with $n_c$ qubits one needs the resonators to be connected with up to $4n_c+1$ tunable couplers. In Fig.~\ref{fig:codesign}, we show sketches of such hardware co-design implementations for unit cells with $n_c\in \{2, 3, 4\}$. Here, resonators are depicted as elongated octagonal red elements positioned between four unit cells (gray) and connect to all of the data qubits they act on (black dots) as well as one internal syndrome qubit (red dot). Note that all couplers on both sides of a flip-chip (thin/thick lines) can be implemented without breaking planarity. Also, note that at least some of the couplers must be connected capacitively, depending on which side of the flip-chip qubits and resonator elements are placed.

% \subsection{Connectivity/Planarity}
% Examples

% \subsection{Stabilizer measurements}

% The derivation of fermionic-to-qubit mappings allow us to protect the underlying quantum algorithm from quantum noise that occurs during its execution by using quantum error-correcting codes. 
% The use of quantum error-correcting codes are twofold: mitigating or correcting errors. 

% TALK ABOUT ERROR MITIGATION WITH QEC and ``reduce the error rate at the cost of more circuit executions as well as small overhead in circuit depth and width''

% BENEFITS OF QEC FOR ERROR CORRECTION BUT THE PROBLEMATIC OF UNKNOWN STRATEGY AND TIME OVERHEAD

% However, it is not clear the optimum (or sub-optimum) strategy of combining syndrome extractions XXX and the implementation of fermionic gates.  

% See arXiv:2306.16608 and arXiv:2309.11673.

% \subsection{Projective measurements}

% Passive QEC? arXiv:2307.09512

\section{Conclusions}
\label{sec:Conclusions}
We have investigated the feasibility of using high-distance fermion-to-qubit encodings for fermionic simulation and identified a number of promising candidates which manage to strike the right balance between multiple seemingly irreconcilable requirements, such as high code distance, low logical operator weights, low stabilizer weights, low fermion-to-qubit ratio, (bi-)planarity and low qubit connectivity. Naturally, as distance increases, so must the weights of logical operators and we observe that they can be pushed down to roughly one or two higher than $d$. Less straightforwardly, we have equally observed an increase in the weights of stabilizers at a pace similar to that of logical operators. Since, a priori, stabilizer weights are not bounded by the distance, it was possible to limit their growth by using encodings with smaller, triangular edge-loops. In the same way, we found that higher qubit-to-mode ratios reduce operator and stabilizer weights of encodings by allowing for more freedom in resolving commutation relations. One interesting finding is that, at higher distances, encodings with additional diagonal edges performed better than those with only horizontal and vertical ones. One could certainly attempt adding unique ever-higher-neighbor edges to encodings, which could potentially lead to even better performance. Similarly, it would be of interest to study encodings defined on other lattices, e.g. triangular, honeycomb or cubic. An alternative idea for generating encodings is to  map Hamiltonian terms (such as hoppings and density-density terms) directly to Pauli operators, thus skipping the intermediate step of defining edges and vertices. This introduces additional freedom in generating encodings, but also comes with more relations that need to be satisfied. For example, for a square lattice, one needs to reconcile the commutation relations of eight hopping operators rather than four edge operators. We have investigated this particular type of encodings using our algorithms and have identified examples of such, but we found that they did not exhibit any favourable properties compared to the standard approach. This is corroborating the findings of Ref.~\cite{chien2022optimizing}, where the authors arrived at a similar conclusion. Finally, we also found that with growing distance, the necessary hardware graph connectivity structure becomes increasingly involved and less amenable to realistic realisations in e.g. superconducting qubit architectures. Of course, this is less of an issue for platforms with all-to-all connectivity.

The growing weights of logical operators as well as stabilizers lead to increasing gate counts and circuit depths, which in turn leads to additional errors being generated during the circuit execution. This overhead has to be put in relation with any error-mitigating or error-correcting capacities of these encodings. Due to a lack of some transversal gates and no strategy for implementing magic state injection, none of the local fermion-to-qubit encodings can perform fault-tolerant universal quantum computations. This means that at a given error rate, eventually errors take over. In Ref.~\cite{chien2023simulating} the authors compared these two effects for a set of encodings and found that for a relatively high noise rate ($1\%$ depolarising noise),  low-distance local encodings performed better than both the Jordan-Wigner transformation (which notably lacks any stabilizers) and high-distance ones. As the authors of Ref.~\cite{chien2023simulating} state, this result might strongly depend on the noise rate and we expect high-distance encodings to improve their relative performance as it is decreased. It would thus be interesting to repeat this comparison also for lower noise rates and possibly with more involved error models. 

Going forward, one open question is how one can keep systematically increasing the distances starting from known local fermion-to-qubit encodings. For some quantum error correcting codes, such as the surface code, increasing the number of qubits increases the code distance. For local fermion-to-qubit encodings, this instead increases the size of the fermionic system whilst keeping the distance constant. In principle, one could attempt to concatenate a fermionic encoding with a quantum error-correcting code in order to be able to increase the distance, but that will likely result in a very large qubit overhead. To the best of our knowledge, the only attempt at this was presented in Ref.~\cite{landahl2023logical}, where Majorana operators are introduced as defects within a surface code. Compared to our encodings, these codes come with a significant resource overhead in terms of both the number of qubits and logical operator weights. 

Finally, it would be of interest to perform similar investigations into high-distance encodings for models with bosonic or anyonic degrees of freedom, which is something we are planning to do in future studies. 

\bibliography{main}% Produces the bibliography via BibTeX.

\providecommand{\noopsort}[1]{}\providecommand{\singleletter}[1]{#1}%
\begin{thebibliography}{38}%
\makeatletter
\providecommand \@ifxundefined [1]{%
 \@ifx{#1\undefined}
}%
\providecommand \@ifnum [1]{%
 \ifnum #1\expandafter \@firstoftwo
 \else \expandafter \@secondoftwo
 \fi
}%
\providecommand \@ifx [1]{%
 \ifx #1\expandafter \@firstoftwo
 \else \expandafter \@secondoftwo
 \fi
}%
\providecommand \natexlab [1]{#1}%
\providecommand \enquote  [1]{``#1''}%
\providecommand \bibnamefont  [1]{#1}%
\providecommand \bibfnamefont [1]{#1}%
\providecommand \citenamefont [1]{#1}%
\providecommand \href@noop [0]{\@secondoftwo}%
\providecommand \href [0]{\begingroup \@sanitize@url \@href}%
\providecommand \@href[1]{\@@startlink{#1}\@@href}%
\providecommand \@@href[1]{\endgroup#1\@@endlink}%
\providecommand \@sanitize@url [0]{\catcode `\\12\catcode `\$12\catcode
  `\&12\catcode `\#12\catcode `\^12\catcode `\_12\catcode `\%12\relax}%
\providecommand \@@startlink[1]{}%
\providecommand \@@endlink[0]{}%
\providecommand \url  [0]{\begingroup\@sanitize@url \@url }%
\providecommand \@url [1]{\endgroup\@href {#1}{\urlprefix }}%
\providecommand \urlprefix  [0]{URL }%
\providecommand \Eprint [0]{\href }%
\providecommand \doibase [0]{https://doi.org/}%
\providecommand \selectlanguage [0]{\@gobble}%
\providecommand \bibinfo  [0]{\@secondoftwo}%
\providecommand \bibfield  [0]{\@secondoftwo}%
\providecommand \translation [1]{[#1]}%
\providecommand \BibitemOpen [0]{}%
\providecommand \bibitemStop [0]{}%
\providecommand \bibitemNoStop [0]{.\EOS\space}%
\providecommand \EOS [0]{\spacefactor3000\relax}%
\providecommand \BibitemShut  [1]{\csname bibitem#1\endcsname}%
\let\auto@bib@innerbib\@empty
%</preamble>
\bibitem [{\citenamefont {Dalzell}\ \emph {et~al.}(2023)\citenamefont
  {Dalzell}, \citenamefont {McArdle}, \citenamefont {Berta}, \citenamefont
  {Bienias}, \citenamefont {Chen}, \citenamefont {Gilyén}, \citenamefont
  {Hann}, \citenamefont {Kastoryano}, \citenamefont {Khabiboulline},
  \citenamefont {Kubica}, \citenamefont {Salton}, \citenamefont {Wang},\ and\
  \citenamefont {Brandão}}]{Dalzell2023}%
  \BibitemOpen
  \bibfield  {author} {\bibinfo {author} {\bibfnamefont {A.~M.}\ \bibnamefont
  {Dalzell}}, \bibinfo {author} {\bibfnamefont {S.}~\bibnamefont {McArdle}},
  \bibinfo {author} {\bibfnamefont {M.}~\bibnamefont {Berta}}, \bibinfo
  {author} {\bibfnamefont {P.}~\bibnamefont {Bienias}}, \bibinfo {author}
  {\bibfnamefont {C.-F.}\ \bibnamefont {Chen}}, \bibinfo {author}
  {\bibfnamefont {A.}~\bibnamefont {Gilyén}}, \bibinfo {author} {\bibfnamefont
  {C.~T.}\ \bibnamefont {Hann}}, \bibinfo {author} {\bibfnamefont {M.~J.}\
  \bibnamefont {Kastoryano}}, \bibinfo {author} {\bibfnamefont {E.~T.}\
  \bibnamefont {Khabiboulline}}, \bibinfo {author} {\bibfnamefont
  {A.}~\bibnamefont {Kubica}}, \bibinfo {author} {\bibfnamefont
  {G.}~\bibnamefont {Salton}}, \bibinfo {author} {\bibfnamefont
  {S.}~\bibnamefont {Wang}},\ and\ \bibinfo {author} {\bibfnamefont {F.~G.
  S.~L.}\ \bibnamefont {Brandão}},\ }\href
  {https://doi.org/10.48550/ARXIV.2310.03011} {\bibinfo {title} {Quantum
  algorithms: A survey of applications and end-to-end complexities}} (\bibinfo
  {year} {2023})\BibitemShut {NoStop}%
\bibitem [{\citenamefont {Cai}\ \emph {et~al.}(2023)\citenamefont {Cai},
  \citenamefont {Babbush}, \citenamefont {Benjamin}, \citenamefont {Endo},
  \citenamefont {Huggins}, \citenamefont {Li}, \citenamefont {McClean},\ and\
  \citenamefont {O’Brien}}]{cai2023quantum}%
  \BibitemOpen
  \bibfield  {author} {\bibinfo {author} {\bibfnamefont {Z.}~\bibnamefont
  {Cai}}, \bibinfo {author} {\bibfnamefont {R.}~\bibnamefont {Babbush}},
  \bibinfo {author} {\bibfnamefont {S.~C.}\ \bibnamefont {Benjamin}}, \bibinfo
  {author} {\bibfnamefont {S.}~\bibnamefont {Endo}}, \bibinfo {author}
  {\bibfnamefont {W.~J.}\ \bibnamefont {Huggins}}, \bibinfo {author}
  {\bibfnamefont {Y.}~\bibnamefont {Li}}, \bibinfo {author} {\bibfnamefont
  {J.~R.}\ \bibnamefont {McClean}},\ and\ \bibinfo {author} {\bibfnamefont
  {T.~E.}\ \bibnamefont {O’Brien}},\ }\bibfield  {title} {\bibinfo {title}
  {Quantum error mitigation},\ }\href@noop {} {\bibfield  {journal} {\bibinfo
  {journal} {Reviews of Modern Physics}\ }\textbf {\bibinfo {volume} {95}},\
  \bibinfo {pages} {045005} (\bibinfo {year} {2023})}\BibitemShut {NoStop}%
\bibitem [{\citenamefont {Haah}(2017)}]{Haah2017}%
  \BibitemOpen
  \bibfield  {author} {\bibinfo {author} {\bibfnamefont {J.}~\bibnamefont
  {Haah}},\ }\bibfield  {title} {\bibinfo {title} {Algebraic methods for
  quantum codes on lattices},\ }\href
  {https://doi.org/10.15446/recolma.v50n2.62214} {\bibfield  {journal}
  {\bibinfo  {journal} {Revista Colombiana de Matemáticas}\ }\textbf {\bibinfo
  {volume} {50}},\ \bibinfo {pages} {299} (\bibinfo {year} {2017})}\BibitemShut
  {NoStop}%
\bibitem [{\citenamefont {Setia}\ \emph {et~al.}(2019)\citenamefont {Setia},
  \citenamefont {Bravyi}, \citenamefont {Mezzacapo},\ and\ \citenamefont
  {Whitfield}}]{setia2019superfast}%
  \BibitemOpen
  \bibfield  {author} {\bibinfo {author} {\bibfnamefont {K.}~\bibnamefont
  {Setia}}, \bibinfo {author} {\bibfnamefont {S.}~\bibnamefont {Bravyi}},
  \bibinfo {author} {\bibfnamefont {A.}~\bibnamefont {Mezzacapo}},\ and\
  \bibinfo {author} {\bibfnamefont {J.~D.}\ \bibnamefont {Whitfield}},\
  }\bibfield  {title} {\bibinfo {title} {Superfast encodings for fermionic
  quantum simulation},\ }\href@noop {} {\bibfield  {journal} {\bibinfo
  {journal} {Physical Review Research}\ }\textbf {\bibinfo {volume} {1}},\
  \bibinfo {pages} {033033} (\bibinfo {year} {2019})}\BibitemShut {NoStop}%
\bibitem [{\citenamefont {Bausch}\ \emph {et~al.}(2020)\citenamefont {Bausch},
  \citenamefont {Cubitt}, \citenamefont {Derby},\ and\ \citenamefont
  {Klassen}}]{Bausch2020}%
  \BibitemOpen
  \bibfield  {author} {\bibinfo {author} {\bibfnamefont {J.}~\bibnamefont
  {Bausch}}, \bibinfo {author} {\bibfnamefont {T.}~\bibnamefont {Cubitt}},
  \bibinfo {author} {\bibfnamefont {C.}~\bibnamefont {Derby}},\ and\ \bibinfo
  {author} {\bibfnamefont {J.}~\bibnamefont {Klassen}},\ }\href
  {https://doi.org/10.48550/ARXIV.2003.07125} {\bibinfo {title} {Mitigating
  errors in local fermionic encodings}} (\bibinfo {year} {2020})\BibitemShut
  {NoStop}%
\bibitem [{\citenamefont {Tantivasadakarn}(2020)}]{Tantivasadakarn2020}%
  \BibitemOpen
  \bibfield  {author} {\bibinfo {author} {\bibfnamefont {N.}~\bibnamefont
  {Tantivasadakarn}},\ }\bibfield  {title} {\bibinfo {title} {Jordan-wigner
  dualities for translation-invariant hamiltonians in any dimension: Emergent
  fermions in fracton topological order},\ }\href
  {https://doi.org/10.1103/physrevresearch.2.023353} {\bibfield  {journal}
  {\bibinfo  {journal} {Physical Review Research}\ }\textbf {\bibinfo {volume}
  {2}},\ \bibinfo {pages} {023353} (\bibinfo {year} {2020})}\BibitemShut
  {NoStop}%
\bibitem [{\citenamefont {Chen}\ \emph {et~al.}(2022)\citenamefont {Chen},
  \citenamefont {Gorshkov},\ and\ \citenamefont {Xu}}]{chen2022error}%
  \BibitemOpen
  \bibfield  {author} {\bibinfo {author} {\bibfnamefont {Y.-A.}\ \bibnamefont
  {Chen}}, \bibinfo {author} {\bibfnamefont {A.~V.}\ \bibnamefont {Gorshkov}},\
  and\ \bibinfo {author} {\bibfnamefont {Y.}~\bibnamefont {Xu}},\ }\bibfield
  {title} {\bibinfo {title} {Error-correcting codes for fermionic quantum
  simulation},\ }\href@noop {} {\bibfield  {journal} {\bibinfo  {journal}
  {arXiv preprint arXiv:2210.08411}\ } (\bibinfo {year} {2022})}\BibitemShut
  {NoStop}%
\bibitem [{\citenamefont {Arovas}\ \emph {et~al.}(2022)\citenamefont {Arovas},
  \citenamefont {Berg}, \citenamefont {Kivelson},\ and\ \citenamefont
  {Raghu}}]{arovas2022hubbard}%
  \BibitemOpen
  \bibfield  {author} {\bibinfo {author} {\bibfnamefont {D.~P.}\ \bibnamefont
  {Arovas}}, \bibinfo {author} {\bibfnamefont {E.}~\bibnamefont {Berg}},
  \bibinfo {author} {\bibfnamefont {S.~A.}\ \bibnamefont {Kivelson}},\ and\
  \bibinfo {author} {\bibfnamefont {S.}~\bibnamefont {Raghu}},\ }\bibfield
  {title} {\bibinfo {title} {The hubbard model},\ }\href@noop {} {\bibfield
  {journal} {\bibinfo  {journal} {Annual review of condensed matter physics}\
  }\textbf {\bibinfo {volume} {13}},\ \bibinfo {pages} {239} (\bibinfo {year}
  {2022})}\BibitemShut {NoStop}%
\bibitem [{\citenamefont {Qin}\ \emph {et~al.}(2022)\citenamefont {Qin},
  \citenamefont {Sch{\"a}fer}, \citenamefont {Andergassen}, \citenamefont
  {Corboz},\ and\ \citenamefont {Gull}}]{qin2022hubbard}%
  \BibitemOpen
  \bibfield  {author} {\bibinfo {author} {\bibfnamefont {M.}~\bibnamefont
  {Qin}}, \bibinfo {author} {\bibfnamefont {T.}~\bibnamefont {Sch{\"a}fer}},
  \bibinfo {author} {\bibfnamefont {S.}~\bibnamefont {Andergassen}}, \bibinfo
  {author} {\bibfnamefont {P.}~\bibnamefont {Corboz}},\ and\ \bibinfo {author}
  {\bibfnamefont {E.}~\bibnamefont {Gull}},\ }\bibfield  {title} {\bibinfo
  {title} {The hubbard model: A computational perspective},\ }\href@noop {}
  {\bibfield  {journal} {\bibinfo  {journal} {Annual Review of Condensed Matter
  Physics}\ }\textbf {\bibinfo {volume} {13}},\ \bibinfo {pages} {275}
  (\bibinfo {year} {2022})}\BibitemShut {NoStop}%
\bibitem [{\citenamefont {Jordan}\ and\ \citenamefont
  {Wigner}(1928)}]{Jordan1928}%
  \BibitemOpen
  \bibfield  {author} {\bibinfo {author} {\bibfnamefont {P.}~\bibnamefont
  {Jordan}}\ and\ \bibinfo {author} {\bibfnamefont {E.}~\bibnamefont
  {Wigner}},\ }\bibfield  {title} {\bibinfo {title} {\"uber das paulische
  \"aquivalenzverbot},\ }\href {https://doi.org/10.1007/bf01331938} {\bibfield
  {journal} {\bibinfo  {journal} {Zeitschrift f\"ur Physik}\ }\textbf {\bibinfo
  {volume} {47}},\ \bibinfo {pages} {631} (\bibinfo {year} {1928})}\BibitemShut
  {NoStop}%
\bibitem [{\citenamefont {Bravyi}\ \emph {et~al.}(2017)\citenamefont {Bravyi},
  \citenamefont {Gambetta}, \citenamefont {Mezzacapo},\ and\ \citenamefont
  {Temme}}]{Bravyi2017}%
  \BibitemOpen
  \bibfield  {author} {\bibinfo {author} {\bibfnamefont {S.}~\bibnamefont
  {Bravyi}}, \bibinfo {author} {\bibfnamefont {J.~M.}\ \bibnamefont
  {Gambetta}}, \bibinfo {author} {\bibfnamefont {A.}~\bibnamefont
  {Mezzacapo}},\ and\ \bibinfo {author} {\bibfnamefont {K.}~\bibnamefont
  {Temme}},\ }\href {https://doi.org/10.48550/ARXIV.1701.08213} {\bibinfo
  {title} {Tapering off qubits to simulate fermionic hamiltonians}} (\bibinfo
  {year} {2017})\BibitemShut {NoStop}%
\bibitem [{\citenamefont {Havlíček}\ \emph {et~al.}(2017)\citenamefont
  {Havlíček}, \citenamefont {Troyer},\ and\ \citenamefont
  {Whitfield}}]{Havlicek2017}%
  \BibitemOpen
  \bibfield  {author} {\bibinfo {author} {\bibfnamefont {V.}~\bibnamefont
  {Havlíček}}, \bibinfo {author} {\bibfnamefont {M.}~\bibnamefont {Troyer}},\
  and\ \bibinfo {author} {\bibfnamefont {J.~D.}\ \bibnamefont {Whitfield}},\
  }\bibfield  {title} {\bibinfo {title} {Operator locality in the quantum
  simulation of fermionic models},\ }\href
  {https://doi.org/10.1103/physreva.95.032332} {\bibfield  {journal} {\bibinfo
  {journal} {Physical Review A}\ }\textbf {\bibinfo {volume} {95}},\ \bibinfo
  {pages} {032332} (\bibinfo {year} {2017})}\BibitemShut {NoStop}%
\bibitem [{\citenamefont {Steudtner}\ and\ \citenamefont
  {Wehner}(2018)}]{Steudtner2018}%
  \BibitemOpen
  \bibfield  {author} {\bibinfo {author} {\bibfnamefont {M.}~\bibnamefont
  {Steudtner}}\ and\ \bibinfo {author} {\bibfnamefont {S.}~\bibnamefont
  {Wehner}},\ }\bibfield  {title} {\bibinfo {title} {Fermion-to-qubit mappings
  with varying resource requirements for quantum simulation},\ }\href
  {https://doi.org/10.1088/1367-2630/aac54f} {\bibfield  {journal} {\bibinfo
  {journal} {New Journal of Physics}\ }\textbf {\bibinfo {volume} {20}},\
  \bibinfo {pages} {063010} (\bibinfo {year} {2018})}\BibitemShut {NoStop}%
\bibitem [{\citenamefont {Jiang}\ \emph {et~al.}(2019)\citenamefont {Jiang},
  \citenamefont {McClean}, \citenamefont {Babbush},\ and\ \citenamefont
  {Neven}}]{jiang2019majorana}%
  \BibitemOpen
  \bibfield  {author} {\bibinfo {author} {\bibfnamefont {Z.}~\bibnamefont
  {Jiang}}, \bibinfo {author} {\bibfnamefont {J.}~\bibnamefont {McClean}},
  \bibinfo {author} {\bibfnamefont {R.}~\bibnamefont {Babbush}},\ and\ \bibinfo
  {author} {\bibfnamefont {H.}~\bibnamefont {Neven}},\ }\bibfield  {title}
  {\bibinfo {title} {Majorana loop stabilizer codes for error mitigation in
  fermionic quantum simulations},\ }\href
  {https://doi.org/10.1103/PhysRevApplied.12.064041} {\bibfield  {journal}
  {\bibinfo  {journal} {Phys. Rev. Appl.}\ }\textbf {\bibinfo {volume} {12}},\
  \bibinfo {pages} {064041} (\bibinfo {year} {2019})}\BibitemShut {NoStop}%
\bibitem [{\citenamefont {Harrison}\ \emph {et~al.}(2022)\citenamefont
  {Harrison}, \citenamefont {Nelson}, \citenamefont {Adamiak},\ and\
  \citenamefont {Whitfield}}]{Harrison2022}%
  \BibitemOpen
  \bibfield  {author} {\bibinfo {author} {\bibfnamefont {B.}~\bibnamefont
  {Harrison}}, \bibinfo {author} {\bibfnamefont {D.}~\bibnamefont {Nelson}},
  \bibinfo {author} {\bibfnamefont {D.}~\bibnamefont {Adamiak}},\ and\ \bibinfo
  {author} {\bibfnamefont {J.}~\bibnamefont {Whitfield}},\ }\href
  {https://doi.org/10.48550/ARXIV.2211.04501} {\bibinfo {title} {Reducing the
  qubit requirement of jordan-wigner encodings of $n$-mode, $k$-fermion systems
  from $n$ to $\lceil \log_2 {N \choose K} \rceil$}} (\bibinfo {year}
  {2022})\BibitemShut {NoStop}%
\bibitem [{\citenamefont {Kirby}\ \emph {et~al.}(2022)\citenamefont {Kirby},
  \citenamefont {Fuller}, \citenamefont {Hadfield},\ and\ \citenamefont
  {Mezzacapo}}]{Kirby2022}%
  \BibitemOpen
  \bibfield  {author} {\bibinfo {author} {\bibfnamefont {W.}~\bibnamefont
  {Kirby}}, \bibinfo {author} {\bibfnamefont {B.}~\bibnamefont {Fuller}},
  \bibinfo {author} {\bibfnamefont {C.}~\bibnamefont {Hadfield}},\ and\
  \bibinfo {author} {\bibfnamefont {A.}~\bibnamefont {Mezzacapo}},\ }\bibfield
  {title} {\bibinfo {title} {Second-quantized fermionic operators with
  polylogarithmic qubit and gate complexity},\ }\href
  {https://doi.org/10.1103/prxquantum.3.020351} {\bibfield  {journal} {\bibinfo
   {journal} {PRX Quantum}\ }\textbf {\bibinfo {volume} {3}},\ \bibinfo {pages}
  {020351} (\bibinfo {year} {2022})}\BibitemShut {NoStop}%
\bibitem [{\citenamefont {Chien}\ and\ \citenamefont
  {Klassen}(2022)}]{chien2022optimizing}%
  \BibitemOpen
  \bibfield  {author} {\bibinfo {author} {\bibfnamefont {R.~W.}\ \bibnamefont
  {Chien}}\ and\ \bibinfo {author} {\bibfnamefont {J.}~\bibnamefont
  {Klassen}},\ }\bibfield  {title} {\bibinfo {title} {Optimizing fermionic
  encodings for both hamiltonian and hardware},\ }\href@noop {} {\bibfield
  {journal} {\bibinfo  {journal} {arXiv preprint arXiv:2210.05652}\ } (\bibinfo
  {year} {2022})}\BibitemShut {NoStop}%
\bibitem [{\citenamefont {Chiew}\ and\ \citenamefont
  {Strelchuk}(2023)}]{Chiew2023}%
  \BibitemOpen
  \bibfield  {author} {\bibinfo {author} {\bibfnamefont {M.}~\bibnamefont
  {Chiew}}\ and\ \bibinfo {author} {\bibfnamefont {S.}~\bibnamefont
  {Strelchuk}},\ }\bibfield  {title} {\bibinfo {title} {Discovering optimal
  fermion-qubit mappings through algorithmic enumeration},\ }\href
  {https://doi.org/10.22331/q-2023-10-18-1145} {\bibfield  {journal} {\bibinfo
  {journal} {Quantum}\ }\textbf {\bibinfo {volume} {7}},\ \bibinfo {pages}
  {1145} (\bibinfo {year} {2023})}\BibitemShut {NoStop}%
\bibitem [{\citenamefont {Algaba}\ \emph {et~al.}(2023)\citenamefont {Algaba},
  \citenamefont {Sriluckshmy}, \citenamefont {Leib},\ and\ \citenamefont
  {Simkovic~IV}}]{algaba2023low}%
  \BibitemOpen
  \bibfield  {author} {\bibinfo {author} {\bibfnamefont {M.~G.}\ \bibnamefont
  {Algaba}}, \bibinfo {author} {\bibfnamefont {P.}~\bibnamefont {Sriluckshmy}},
  \bibinfo {author} {\bibfnamefont {M.}~\bibnamefont {Leib}},\ and\ \bibinfo
  {author} {\bibfnamefont {F.}~\bibnamefont {Simkovic~IV}},\ }\bibfield
  {title} {\bibinfo {title} {Low-depth simulations of fermionic systems on
  square-grid quantum hardware},\ }\href@noop {} {\bibfield  {journal}
  {\bibinfo  {journal} {arXiv preprint arXiv:2302.01862}\ } (\bibinfo {year}
  {2023})}\BibitemShut {NoStop}%
\bibitem [{\citenamefont {Verstraete}\ and\ \citenamefont
  {Cirac}(2005)}]{verstraete2005mapping}%
  \BibitemOpen
  \bibfield  {author} {\bibinfo {author} {\bibfnamefont {F.}~\bibnamefont
  {Verstraete}}\ and\ \bibinfo {author} {\bibfnamefont {J.~I.}\ \bibnamefont
  {Cirac}},\ }\bibfield  {title} {\bibinfo {title} {Mapping local hamiltonians
  of fermions to local hamiltonians of spins},\ }\href@noop {} {\bibfield
  {journal} {\bibinfo  {journal} {Journal of Statistical Mechanics: Theory and
  Experiment}\ }\textbf {\bibinfo {volume} {2005}},\ \bibinfo {pages} {P09012}
  (\bibinfo {year} {2005})}\BibitemShut {NoStop}%
\bibitem [{\citenamefont {Derby}\ \emph {et~al.}(2021)\citenamefont {Derby},
  \citenamefont {Klassen}, \citenamefont {Bausch},\ and\ \citenamefont
  {Cubitt}}]{derby2021compact}%
  \BibitemOpen
  \bibfield  {author} {\bibinfo {author} {\bibfnamefont {C.}~\bibnamefont
  {Derby}}, \bibinfo {author} {\bibfnamefont {J.}~\bibnamefont {Klassen}},
  \bibinfo {author} {\bibfnamefont {J.}~\bibnamefont {Bausch}},\ and\ \bibinfo
  {author} {\bibfnamefont {T.}~\bibnamefont {Cubitt}},\ }\bibfield  {title}
  {\bibinfo {title} {Compact fermion to qubit mappings},\ }\href
  {https://doi.org/10.1103/PhysRevB.104.035118} {\bibfield  {journal} {\bibinfo
   {journal} {Phys. Rev. B}\ }\textbf {\bibinfo {volume} {104}},\ \bibinfo
  {pages} {035118} (\bibinfo {year} {2021})}\BibitemShut {NoStop}%
\bibitem [{\citenamefont {Ball}(2005)}]{ball2005fermions}%
  \BibitemOpen
  \bibfield  {author} {\bibinfo {author} {\bibfnamefont {R.}~\bibnamefont
  {Ball}},\ }\bibfield  {title} {\bibinfo {title} {Fermions without fermion
  fields},\ }\href@noop {} {\bibfield  {journal} {\bibinfo  {journal} {Physical
  review letters}\ }\textbf {\bibinfo {volume} {95}},\ \bibinfo {pages}
  {176407} (\bibinfo {year} {2005})}\BibitemShut {NoStop}%
\bibitem [{\citenamefont {Steudtner}\ and\ \citenamefont
  {Wehner}(2017)}]{steudtner2017lowering}%
  \BibitemOpen
  \bibfield  {author} {\bibinfo {author} {\bibfnamefont {M.}~\bibnamefont
  {Steudtner}}\ and\ \bibinfo {author} {\bibfnamefont {S.}~\bibnamefont
  {Wehner}},\ }\bibfield  {title} {\bibinfo {title} {Lowering qubit
  requirements for quantum simulations of fermionic systems},\ }\href@noop {}
  {\bibfield  {journal} {\bibinfo  {journal} {arXiv preprint arXiv:1712.07067}\
  } (\bibinfo {year} {2017})}\BibitemShut {NoStop}%
\bibitem [{\citenamefont {Chen}\ \emph {et~al.}(2018)\citenamefont {Chen},
  \citenamefont {Kapustin},\ and\ \citenamefont
  {Radi{\v{c}}evi{\'c}}}]{chen2018exact}%
  \BibitemOpen
  \bibfield  {author} {\bibinfo {author} {\bibfnamefont {Y.-A.}\ \bibnamefont
  {Chen}}, \bibinfo {author} {\bibfnamefont {A.}~\bibnamefont {Kapustin}},\
  and\ \bibinfo {author} {\bibfnamefont {{\DJ}.}~\bibnamefont
  {Radi{\v{c}}evi{\'c}}},\ }\bibfield  {title} {\bibinfo {title} {Exact
  bosonization in two spatial dimensions and a new class of lattice gauge
  theories},\ }\href@noop {} {\bibfield  {journal} {\bibinfo  {journal} {Annals
  of Physics}\ }\textbf {\bibinfo {volume} {393}},\ \bibinfo {pages} {234}
  (\bibinfo {year} {2018})}\BibitemShut {NoStop}%
\bibitem [{\citenamefont {Chien}\ \emph {et~al.}(2023)\citenamefont {Chien},
  \citenamefont {Setia}, \citenamefont {Bonet-Monroig}, \citenamefont
  {Steudtner},\ and\ \citenamefont {Whitfield}}]{chien2023simulating}%
  \BibitemOpen
  \bibfield  {author} {\bibinfo {author} {\bibfnamefont {R.~W.}\ \bibnamefont
  {Chien}}, \bibinfo {author} {\bibfnamefont {K.}~\bibnamefont {Setia}},
  \bibinfo {author} {\bibfnamefont {X.}~\bibnamefont {Bonet-Monroig}}, \bibinfo
  {author} {\bibfnamefont {M.}~\bibnamefont {Steudtner}},\ and\ \bibinfo
  {author} {\bibfnamefont {J.~D.}\ \bibnamefont {Whitfield}},\ }\bibfield
  {title} {\bibinfo {title} {Simulating quantum error mitigation in fermionic
  encodings},\ }\href@noop {} {\bibfield  {journal} {\bibinfo  {journal} {arXiv
  preprint arXiv:2303.02270}\ } (\bibinfo {year} {2023})}\BibitemShut {NoStop}%
\bibitem [{\citenamefont {Nielsen}\ \emph {et~al.}(2010)\citenamefont
  {Nielsen}, \citenamefont {Nielsen}, \citenamefont {Chuang},\ and\
  \citenamefont {INDIA}}]{Nielsen2010}%
  \BibitemOpen
  \bibfield  {author} {\bibinfo {author} {\bibfnamefont {M.~A.}\ \bibnamefont
  {Nielsen}}, \bibinfo {author} {\bibfnamefont {M.~A.}\ \bibnamefont
  {Nielsen}}, \bibinfo {author} {\bibfnamefont {I.~L.}\ \bibnamefont
  {Chuang}},\ and\ \bibinfo {author} {\bibfnamefont {C.~A. M. B. R. I. D.
  G.~E.}\ \bibnamefont {INDIA}},\ }\href@noop {} {\emph {\bibinfo {title}
  {Quantum computation and quantum information - 10. ed.}}}\ (\bibinfo
  {publisher} {Cambridge University Press},\ \bibinfo {year} {2010})\ p.\
  \bibinfo {pages} {702}\BibitemShut {NoStop}%
\bibitem [{\citenamefont {Fowler}\ \emph {et~al.}(2012)\citenamefont {Fowler},
  \citenamefont {Mariantoni}, \citenamefont {Martinis},\ and\ \citenamefont
  {Cleland}}]{Fowler2012}%
  \BibitemOpen
  \bibfield  {author} {\bibinfo {author} {\bibfnamefont {A.~G.}\ \bibnamefont
  {Fowler}}, \bibinfo {author} {\bibfnamefont {M.}~\bibnamefont {Mariantoni}},
  \bibinfo {author} {\bibfnamefont {J.~M.}\ \bibnamefont {Martinis}},\ and\
  \bibinfo {author} {\bibfnamefont {A.~N.}\ \bibnamefont {Cleland}},\
  }\bibfield  {title} {\bibinfo {title} {Surface codes: Towards practical
  large-scale quantum computation},\ }\href
  {https://doi.org/10.1103/physreva.86.032324} {\bibfield  {journal} {\bibinfo
  {journal} {Physical Review A}\ }\textbf {\bibinfo {volume} {86}},\ \bibinfo
  {pages} {032324} (\bibinfo {year} {2012})}\BibitemShut {NoStop}%
\bibitem [{\citenamefont {Vardy}(1997)}]{Vardy1997}%
  \BibitemOpen
  \bibfield  {author} {\bibinfo {author} {\bibfnamefont {A.}~\bibnamefont
  {Vardy}},\ }\bibfield  {title} {\bibinfo {title} {The intractability of
  computing the minimum distance of a code},\ }\href
  {https://doi.org/10.1109/18.641542} {\bibfield  {journal} {\bibinfo
  {journal} {IEEE Transactions on Information Theory}\ }\textbf {\bibinfo
  {volume} {43}},\ \bibinfo {pages} {1757} (\bibinfo {year}
  {1997})}\BibitemShut {NoStop}%
\bibitem [{\citenamefont {Kapshikar}\ and\ \citenamefont
  {Kundu}(2023)}]{Kapshikar2023}%
  \BibitemOpen
  \bibfield  {author} {\bibinfo {author} {\bibfnamefont {U.}~\bibnamefont
  {Kapshikar}}\ and\ \bibinfo {author} {\bibfnamefont {S.}~\bibnamefont
  {Kundu}},\ }\bibfield  {title} {\bibinfo {title} {On the hardness of the
  minimum distance problem of quantum codes},\ }\href
  {https://doi.org/10.1109/tit.2023.3286870} {\bibfield  {journal} {\bibinfo
  {journal} {IEEE Transactions on Information Theory}\ }\textbf {\bibinfo
  {volume} {69}},\ \bibinfo {pages} {6293} (\bibinfo {year}
  {2023})}\BibitemShut {NoStop}%
\bibitem [{\citenamefont {Aaronson}\ and\ \citenamefont
  {Gottesman}(2004)}]{Aaronson2004}%
  \BibitemOpen
  \bibfield  {author} {\bibinfo {author} {\bibfnamefont {S.}~\bibnamefont
  {Aaronson}}\ and\ \bibinfo {author} {\bibfnamefont {D.}~\bibnamefont
  {Gottesman}},\ }\bibfield  {title} {\bibinfo {title} {Improved simulation of
  stabilizer circuits},\ }\href {https://doi.org/10.1103/physreva.70.052328}
  {\bibfield  {journal} {\bibinfo  {journal} {Physical Review A}\ }\textbf
  {\bibinfo {volume} {70}},\ \bibinfo {pages} {052328} (\bibinfo {year}
  {2004})}\BibitemShut {NoStop}%
\bibitem [{\citenamefont {Gottesman}(1998{\natexlab{a}})}]{Gottesman_1998}%
  \BibitemOpen
  \bibfield  {author} {\bibinfo {author} {\bibfnamefont {D.}~\bibnamefont
  {Gottesman}},\ }\bibfield  {title} {\bibinfo {title} {Theory of
  fault-tolerant quantum computation},\ }\href
  {https://doi.org/10.1103/PhysRevA.57.127} {\bibfield  {journal} {\bibinfo
  {journal} {Phys. Rev. A}\ }\textbf {\bibinfo {volume} {57}},\ \bibinfo
  {pages} {127} (\bibinfo {year} {1998}{\natexlab{a}})}\BibitemShut {NoStop}%
\bibitem [{\citenamefont
  {Gottesman}(1998{\natexlab{b}})}]{gottesman1998heisenberg}%
  \BibitemOpen
  \bibfield  {author} {\bibinfo {author} {\bibfnamefont {D.}~\bibnamefont
  {Gottesman}},\ }\href@noop {} {\bibinfo {title} {The heisenberg
  representation of quantum computers}} (\bibinfo {year}
  {1998}{\natexlab{b}}),\ \Eprint {https://arxiv.org/abs/quant-ph/9807006}
  {arXiv:quant-ph/9807006 [quant-ph]} \BibitemShut {NoStop}%
\bibitem [{\citenamefont {Chen}\ and\ \citenamefont
  {Xu}(2023{\natexlab{a}})}]{chen2023equivalence}%
  \BibitemOpen
  \bibfield  {author} {\bibinfo {author} {\bibfnamefont {Y.-A.}\ \bibnamefont
  {Chen}}\ and\ \bibinfo {author} {\bibfnamefont {Y.}~\bibnamefont {Xu}},\
  }\bibfield  {title} {\bibinfo {title} {Equivalence between fermion-to-qubit
  mappings in two spatial dimensions},\ }\href
  {https://doi.org/10.1103/PRXQuantum.4.010326} {\bibfield  {journal} {\bibinfo
   {journal} {PRX Quantum}\ }\textbf {\bibinfo {volume} {4}},\ \bibinfo {pages}
  {010326} (\bibinfo {year} {2023}{\natexlab{a}})}\BibitemShut {NoStop}%
\bibitem [{\citenamefont {Landahl}\ and\ \citenamefont
  {Morrison}(2023)}]{landahl2023logical}%
  \BibitemOpen
  \bibfield  {author} {\bibinfo {author} {\bibfnamefont {A.~J.}\ \bibnamefont
  {Landahl}}\ and\ \bibinfo {author} {\bibfnamefont {B.~C.~A.}\ \bibnamefont
  {Morrison}},\ }\href@noop {} {\bibinfo {title} {Logical fermions for
  fault-tolerant quantum simulation}} (\bibinfo {year} {2023}),\ \Eprint
  {https://arxiv.org/abs/2110.10280} {arXiv:2110.10280 [quant-ph]} \BibitemShut
  {NoStop}%
\bibitem [{\citenamefont {Chen}\ and\ \citenamefont
  {Xu}(2023{\natexlab{b}})}]{Chen_2023}%
  \BibitemOpen
  \bibfield  {author} {\bibinfo {author} {\bibfnamefont {Y.-A.}\ \bibnamefont
  {Chen}}\ and\ \bibinfo {author} {\bibfnamefont {Y.}~\bibnamefont {Xu}},\
  }\bibfield  {title} {\bibinfo {title} {Equivalence between fermion-to-qubit
  mappings in two spatial dimensions},\ }\href
  {https://doi.org/10.1103/PRXQuantum.4.010326} {\bibfield  {journal} {\bibinfo
   {journal} {PRX Quantum}\ }\textbf {\bibinfo {volume} {4}},\ \bibinfo {pages}
  {010326} (\bibinfo {year} {2023}{\natexlab{b}})}\BibitemShut {NoStop}%
\bibitem [{\citenamefont {McClean}\ \emph {et~al.}(2020)\citenamefont
  {McClean}, \citenamefont {Jiang}, \citenamefont {Rubin}, \citenamefont
  {Babbush},\ and\ \citenamefont {Neven}}]{mcclean2020decoding}%
  \BibitemOpen
  \bibfield  {author} {\bibinfo {author} {\bibfnamefont {J.~R.}\ \bibnamefont
  {McClean}}, \bibinfo {author} {\bibfnamefont {Z.}~\bibnamefont {Jiang}},
  \bibinfo {author} {\bibfnamefont {N.~C.}\ \bibnamefont {Rubin}}, \bibinfo
  {author} {\bibfnamefont {R.}~\bibnamefont {Babbush}},\ and\ \bibinfo {author}
  {\bibfnamefont {H.}~\bibnamefont {Neven}},\ }\bibfield  {title} {\bibinfo
  {title} {Decoding quantum errors with subspace expansions},\ }\bibfield
  {journal} {\bibinfo  {journal} {Nature Communications}\ }\textbf {\bibinfo
  {volume} {11}},\ \href {https://doi.org/10.1038/s41467-020-14341-w}
  {10.1038/s41467-020-14341-w} (\bibinfo {year} {2020})\BibitemShut {NoStop}%
\bibitem [{\citenamefont {Field}\ \emph {et~al.}(2023)\citenamefont {Field},
  \citenamefont {Chen}, \citenamefont {Scharmann}, \citenamefont {Sete},
  \citenamefont {Oruc}, \citenamefont {Vu}, \citenamefont {Kosenko},
  \citenamefont {Mutus}, \citenamefont {Poletto},\ and\ \citenamefont
  {Bestwick}}]{field2023modular}%
  \BibitemOpen
  \bibfield  {author} {\bibinfo {author} {\bibfnamefont {M.}~\bibnamefont
  {Field}}, \bibinfo {author} {\bibfnamefont {A.~Q.}\ \bibnamefont {Chen}},
  \bibinfo {author} {\bibfnamefont {B.}~\bibnamefont {Scharmann}}, \bibinfo
  {author} {\bibfnamefont {E.~A.}\ \bibnamefont {Sete}}, \bibinfo {author}
  {\bibfnamefont {F.}~\bibnamefont {Oruc}}, \bibinfo {author} {\bibfnamefont
  {K.}~\bibnamefont {Vu}}, \bibinfo {author} {\bibfnamefont {V.}~\bibnamefont
  {Kosenko}}, \bibinfo {author} {\bibfnamefont {J.~Y.}\ \bibnamefont {Mutus}},
  \bibinfo {author} {\bibfnamefont {S.}~\bibnamefont {Poletto}},\ and\ \bibinfo
  {author} {\bibfnamefont {A.}~\bibnamefont {Bestwick}},\ }\bibfield  {title}
  {\bibinfo {title} {Modular superconducting qubit architecture with a
  multi-chip tunable coupler},\ }\href@noop {} {\bibfield  {journal} {\bibinfo
  {journal} {arXiv preprint arXiv:2308.09240}\ } (\bibinfo {year}
  {2023})}\BibitemShut {NoStop}%
\bibitem [{\citenamefont {Song}\ \emph {et~al.}(2019)\citenamefont {Song},
  \citenamefont {Xu}, \citenamefont {Li}, \citenamefont {Zhang}, \citenamefont
  {Zhang}, \citenamefont {Liu}, \citenamefont {Guo}, \citenamefont {Wang},
  \citenamefont {Ren}, \citenamefont {Hao} \emph
  {et~al.}}]{song2019generation}%
  \BibitemOpen
  \bibfield  {author} {\bibinfo {author} {\bibfnamefont {C.}~\bibnamefont
  {Song}}, \bibinfo {author} {\bibfnamefont {K.}~\bibnamefont {Xu}}, \bibinfo
  {author} {\bibfnamefont {H.}~\bibnamefont {Li}}, \bibinfo {author}
  {\bibfnamefont {Y.-R.}\ \bibnamefont {Zhang}}, \bibinfo {author}
  {\bibfnamefont {X.}~\bibnamefont {Zhang}}, \bibinfo {author} {\bibfnamefont
  {W.}~\bibnamefont {Liu}}, \bibinfo {author} {\bibfnamefont {Q.}~\bibnamefont
  {Guo}}, \bibinfo {author} {\bibfnamefont {Z.}~\bibnamefont {Wang}}, \bibinfo
  {author} {\bibfnamefont {W.}~\bibnamefont {Ren}}, \bibinfo {author}
  {\bibfnamefont {J.}~\bibnamefont {Hao}}, \emph {et~al.},\ }\bibfield  {title}
  {\bibinfo {title} {Generation of multicomponent atomic schr{\"o}dinger cat
  states of up to 20 qubits},\ }\href@noop {} {\bibfield  {journal} {\bibinfo
  {journal} {Science}\ }\textbf {\bibinfo {volume} {365}},\ \bibinfo {pages}
  {574} (\bibinfo {year} {2019})}\BibitemShut {NoStop}%
\end{thebibliography}%

\onecolumngrid
% \newpage
% \clearpage
\appendix

\newpage

\section{Supplementary results}
\label{appendix}
\newcommand{\centered}[1]{\begin{tabular}{c} #1 \end{tabular}}

In Tables \ref{table:enc1}, \ref{table:enc2}, \ref{table:enc3} and \ref{table:enc4} of this Appendix we provide examples of efficient fermion-to-qubit encodings of distances $2\leq d \leq 7$ from Figs. \ref{fig:combined_plot_2345} and \ref{fig:combined_plot_67}. We label the type of each encodings by two letters, b-d and e-h which correspond to the different options presented in Fig. \ref{fig:best_encodings} of the main text. The definitions of logical operators in terms of Pauli strings are given by the Laurent polynomial matrix $\hat{\sigma}$, defined in Eq.~(\ref{eq:sigma}) of Section \ref{sec:Preliminaries}. The number and type of edge and vertex (logical) operators $f_i$ differs between encodings. We fix their order in the matrix $\hat{\sigma}$ to: $V$, $E_x$, $E_y$, ($E_{xy}$), ($E_{\hat{x}y}$) for encodings with one fermionic mode per unit cell (note that $E_{xy}$, $E_{\hat{x}y}$ do not necessarily exist within encodings). For encodings with two modes per unit cell we set the order to $V_1$, $E_{1,x}$, $E_{1,y}$, ($E_{1,xy}$), ($E_{1,\hat{x}y})$ $V_2$, $E_{2,x}$, $E_{2,y}$, ($E_{2,xy}$), ($E_{2,\hat{x}y}$). The stabilizers, hopping and density-density operators can be easily computed from the definitions of the edge and vertex operators and we only provide their weights in the tables. Additionally, we study the necessary connectivity of encodings for implementing the logical operators of the Fermi-Hubbard model excluding/including next-nearest neighbor hopping operators, and we label the two options as NN/NNN. For each of these we show the minimal connectivity of a 3-by-3 unit cell hardware (qubit) graph in the bulk. Here, unit cells are represented by large black circles, the corresponding qubits by black dots and couplers between qubits (capable of executing two-qubit gates) are shown as blue lines. Furthermore, we represent syndrome cells used for measuring stabilizers as small black circles containing single red dots. Whilst one single qubit (with sufficient connectivity) is enough to measure all stabilizers defined between four unit cells, this does not mean that more syndrome (or flag) qubits couldn't be used for this purpose. The necessary coupler connectivity between syndromes and data qubits within unit cells are shown as thin red lines. Finally, for each connectivity structure we indicate whether all couplers between unit cell qubits can be realised in a way that preserves the planarity of the hardware connectivity graph and also provide the maximum connectivity of any given qubit in the unit cell.

% PAGE 1
\begin{table}
% \centering
[ht] \caption{} \label{tab:encodings}
% \begin{tabular}{|c|c|c|c|} 
\begin{tabular}{|@{}c@{}|@{}c@{}|@{}c@{}|@{}c@{}|}
\hline Logical operators & NN connectivity & NNN connectivity & Additional information \\
\hline 
\centered{
\,
$\begin{bmatrix}
1 & x & 1+x \\
0 & x & y \\
0 & 1 & 1 \\
1 & x & 0 
\end{bmatrix}$
\,}
& \centered{\, \includegraphics[width=3cm]{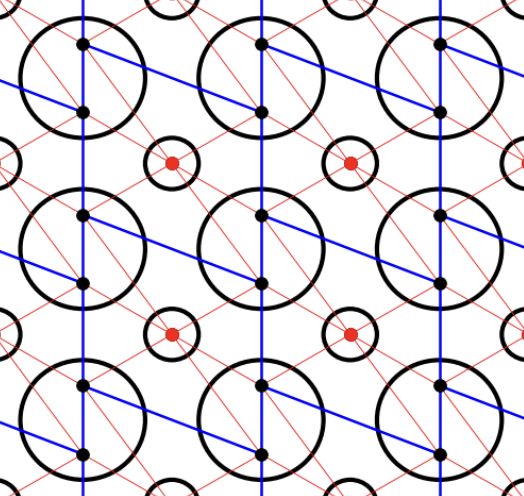} \,} 
& \centered{\, \includegraphics[width=3cm]{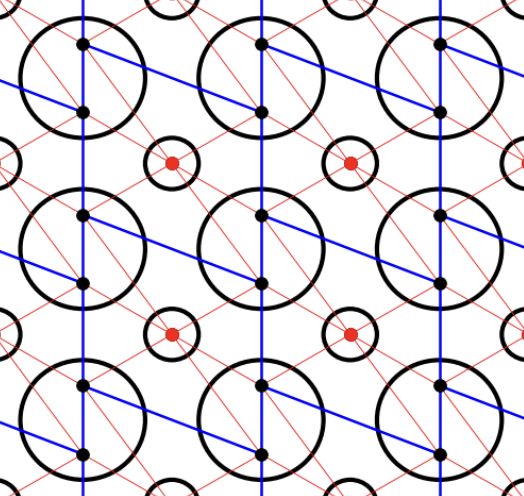} \,} & 
\centered{
type: b, e \\
distance: 2 \\
stabilizers: \{6\} \\
NN hoppings: \{4, 2, 4, 2\} \\
NNN hoppings: \{6, 4, 4, 4\} \\
den-den: 4 \\
planar NN/NNN: yes/yes \\
max. conn. NN/NNN: 3/3
} \\
\hline

\centered{
\,
$\begin{bmatrix}
0 & 0 & 1 & 1+x\\
1 & 1 & 1+y & 1+xy\\
1 & 0 & 1 & 1\\
0 & 0 & 1 & 0\\
0 & 0 & y & xy\\
0 & 1 & 1 & 1
\end{bmatrix}$
\,}
& \centered{\, \includegraphics[width=3cm]{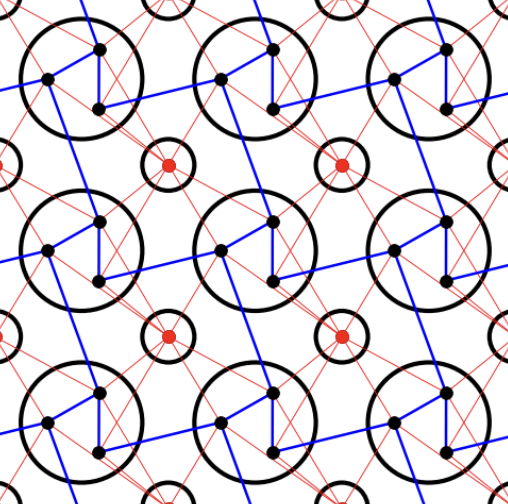} \,} 
& \centered{\, \includegraphics[width=3cm]{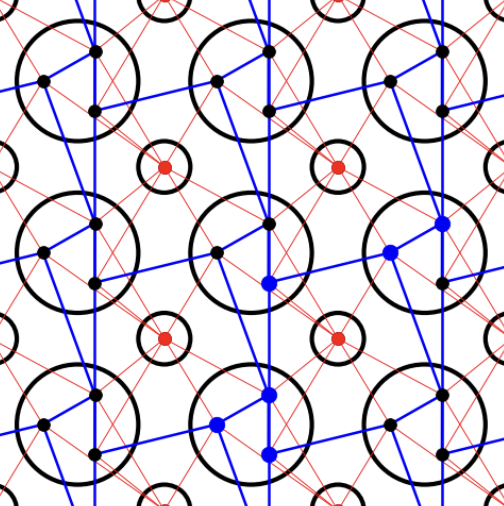} \,} & 
\centered{
type: c, e \\
distance: 2 \\
stabilizers: \{5, 5\} \\
NN hoppings: \{2, 4, 3, 5\} \\
NNN hoppings: \{4, 6, 5, 5\} \\
den-den: 4 \\
planar NN/NNN: yes/yes \\
max. conn. NN/NNN: 3/4
} \\
\hline

\centered{
\,
$\begin{bmatrix}
1   & x    & y\\
1+y & y+xy & 1+y\\
0   & 1+x  & 0\\
0   & 0    & y\\
\end{bmatrix}$
\,}
& \centered{\, \includegraphics[width=3cm]{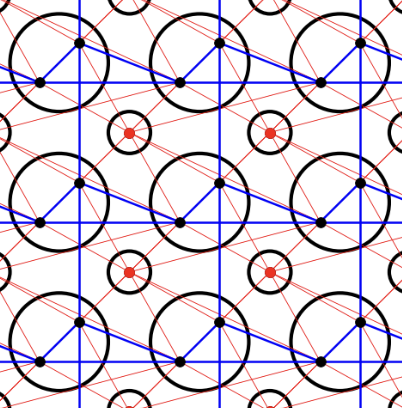} \,} 
& \centered{\, \includegraphics[width=3cm]{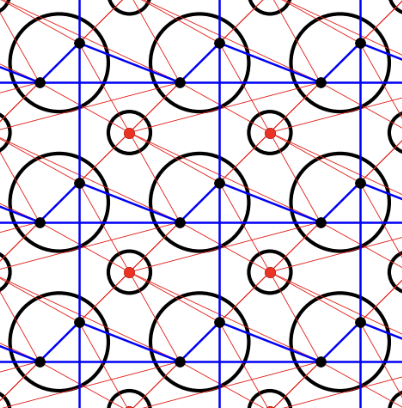} \,} & 
\centered{
type: c, e \\
distance: 3 \\
stabilizers: \{10\} \\
NN hoppings: \{4, 4, 3, 3\} \\
NNN hoppings: \{6, 6, 6, 6\} \\
den-den: 6 \\
planar NN/NNN: no/no \\
max. conn. NN/NNN: 4/4
} \\
\hline

\centered{
\,
$\begin{bmatrix}
1 & x & y & x+xy\\
1 & 1 & y & 1+xy\\
1 & 0 & 1 & 1\\
0 & 0 & 1 & 0\\
0 & x & y & xy\\
0 & 1 & 0 & 1
\end{bmatrix}$
\,}
& \centered{\, \includegraphics[width=3cm]{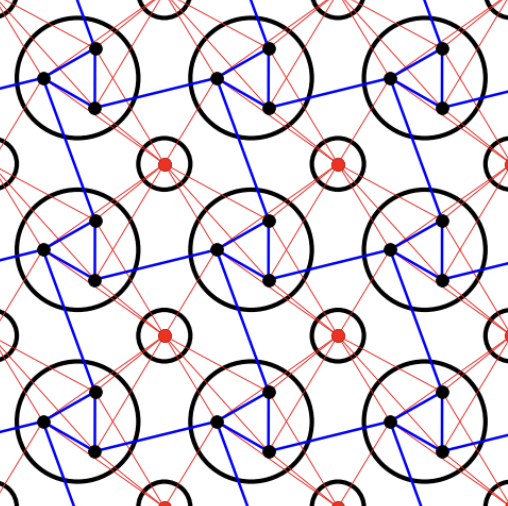} \,} 
& \centered{\, \includegraphics[width=3cm]{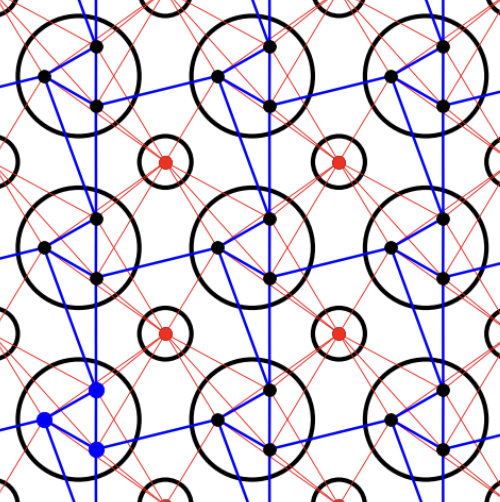} \,} & 
\centered{
type: c, e \\
distance: 3 \\
stabilizers: \{4, 8\} \\
NN hoppings: \{4, 4, 4, 4\} \\
NNN hoppings: \{5, 5, 5, 5\} \\
den-den: 6 \\
planar NN/NNN: yes/yes \\
max. conn. NN/NNN: 4/4
} \\
\hline

\centered{
\,
$\begin{bmatrix}
1 & x & y & xy\\
1 & 1 & 0 & 1+xy\\
1 & 0 & 1+y & 1\\
0 & 0 & 1+y & 0\\
0 & x & 0 & xy\\
0 & 1 & 0 & 1
\end{bmatrix}$
\,}
& \centered{\, \includegraphics[width=3cm]{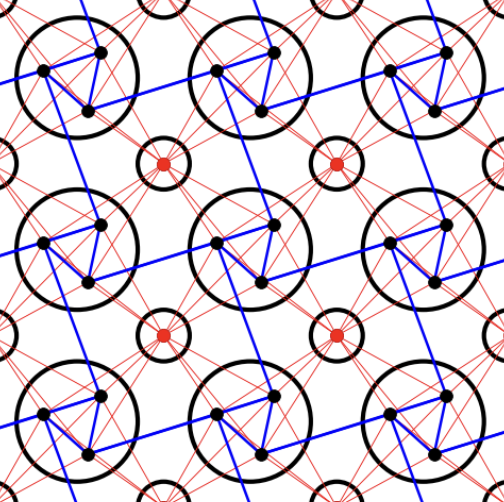} \,} 
& \centered{\, \includegraphics[width=3cm]{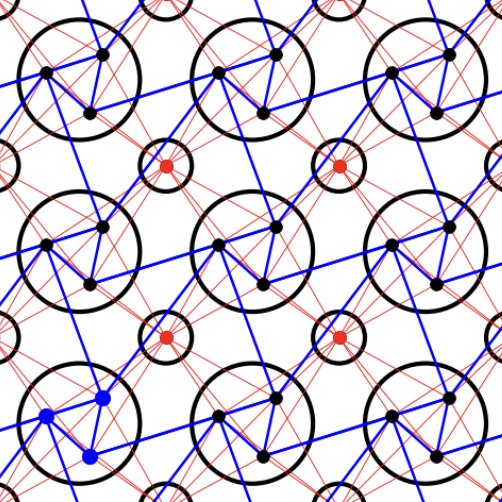} \,} & 
\centered{
type: c, e \\
distance: 3 \\
stabilizers: \{7, 7\} \\
NN hoppings: \{4, 4, 4, 4\} \\
NNN hoppings: \{4, 4, 7, 7\} \\
den-den: 6 \\
planar NN/NNN: yes/yes \\
max. conn. NN/NNN: 4/5
} \\
\hline

\centered{
\,
$\begin{bmatrix}
1 & x & y   & xy   & 1\\
1 & 1 & 0   & 1+xy & y+\bar{x}y\\
1 & 0 & 1+y & 1    & 0\\
0 & 0 & 1+y & 0    & 0\\
0 & x & 0   & xy   & 1\\
0 & 1 & 0   & 1    & \bar{x}+\bar{x}y
\end{bmatrix}$
\,}
& \centered{\, \includegraphics[width=3cm]{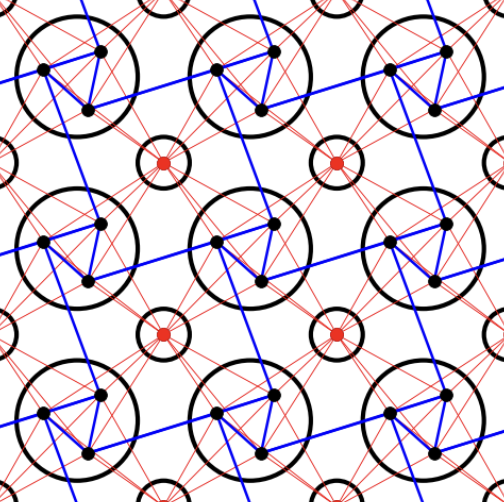} \,} 
& \centered{\, \includegraphics[width=3cm]{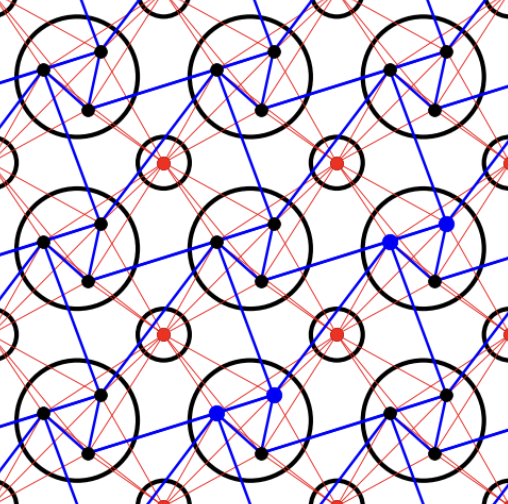} \,} & 
\centered{
type: d, e \\
distance: 3 \\
stabilizers: \{7, 7, 7, 7\} \\
NN hoppings: \{4, 4, 4, 4\} \\
NNN hoppings: \{4, 4, 6, 6\} \\
den-den: 6 \\
planar NN/NNN: yes/yes \\
max. conn. NN/NNN: 4/5
} \\
\hline

\centered{
\,
$\begin{bmatrix}
1 & 1 & 0 & \bar{y} & \bar{y} & xy\\
1 & 0 & 0 & 0       & x       & xy\\
1 & x & 1 & x       & 0       & x\\
0 & 0 & y & 1       & 0       & 1\\
0 & 0 & 1 & 0       & 0       & xy\\
0 & 1 & 0 & 0       & 0       & 0\\
0 & 0 & 0 & 0       & x       & 0\\
0 & 1 & 0 & 0       & 0       & 1
\end{bmatrix}$
\,}
& \centered{\, \includegraphics[width=3cm]{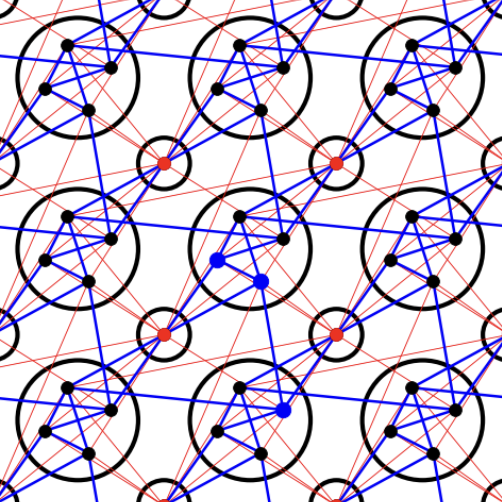} \,} 
& \centered{\, \includegraphics[width=3cm]{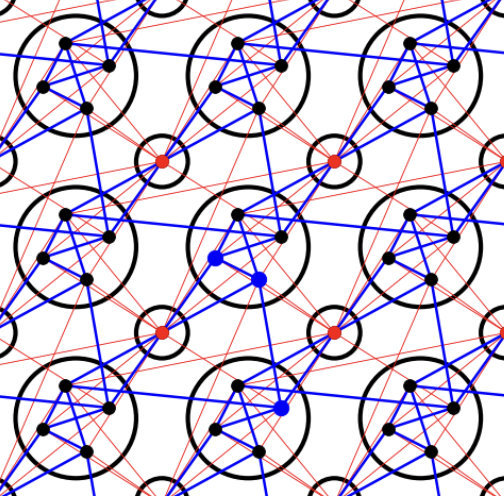} \,} & 
\centered{
type: b, h \\
distance: 3 \\
stabilizers: \{8, 8\} \\
NN hoppings: \{4, 4, 3, 3, 3, 3, 4, 4\} \\
NNN hoppings: \{5, 5, 6, 6, 5, 5, 6, 6\} \\
den-den: 6 \\
planar NN/NNN: yes/yes \\
max. conn. NN/NNN: 4/4
} \\
\hline

\end{tabular}
%   \tabfnt{Note: Bold categories are biologically correct growth models.}
\label{table:enc1}
\end{table}

% PAGE 2
% \hspace{-6cm}

\begin{table}
% \centering
[ht] \caption{} \label{tab:encodings}
% \begin{tabular}{|c|c|c|c|} 
\begin{tabular}{|@{}c@{}|@{}c@{}|@{}c@{}|@{}c@{}|}
\hline Logical operators & NN connectivity & NNN connectivity & Additional information \\
\hline 
\centered{
\,
$\begin{bmatrix}
1+\bar{y} & x+\bar{y}+x\bar{y} & y+\bar{y}   & xy\\
1         & 1                  & 0           & 1+xy\\
1         & 0                  & 1+y         & 1\\
0         & 0                  & 1           & 0\\
0         & x                  & y           & xy\\
0         & 1                  & y           & 1
\end{bmatrix}$
\,}
& \centered{\, \includegraphics[width=3cm]{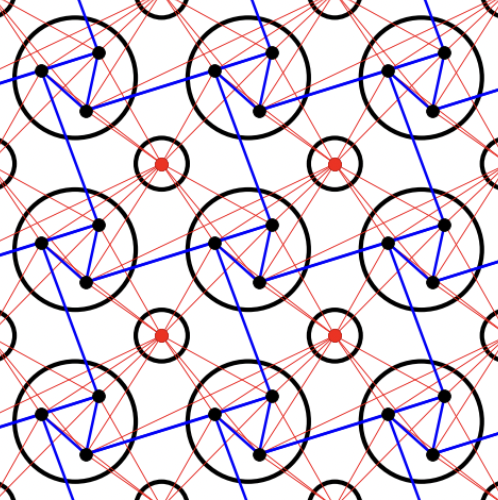} \,} 
& \centered{\, \includegraphics[width=3cm]{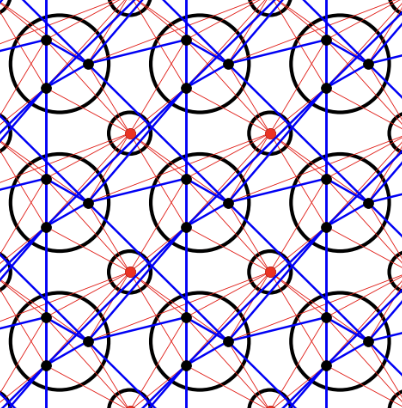} \,} & 
\centered{
type: c, e \\
distance: 4 \\
stabilizers: \{7, 9\} \\
NN hoppings: \{5, 5, 5, 5\} \\
NNN hoppings: \{5, 5, 6, 6\} \\
den-den: 8 \\
planar NN/NNN: yes/no \\
max. conn. NN/NNN: 4/6
} \\
\hline

\centered{
\,
$\begin{bmatrix}
1+x & 0 & 1+x & x+xy\\
1   & 1 & 0   & xy\\
1   & 0 & y   & 1+xy\\
0   & x & 0   & 0\\
0   & x & y   & xy\\
0   & x & 1   & 1
\end{bmatrix}$
\,}
& \centered{\, \includegraphics[width=3cm]{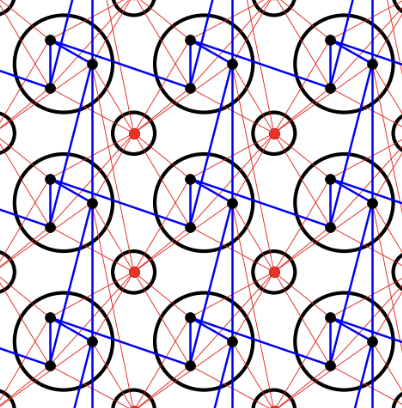} \,} 
& \centered{\, \includegraphics[width=3cm]{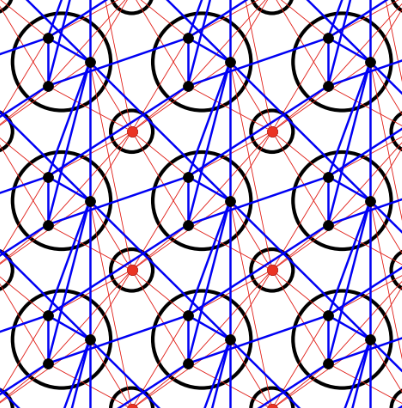} \,} & 
\centered{
type: c, e \\
distance: 4 \\
stabilizers: \{7, 7\} \\
NN hoppings: \{5, 5, 4, 6\} \\
NNN hoppings: \{6, 4, 5, 7\} \\
den-den: 8 \\
planar NN/NNN: no/no \\
max. conn. NN/NNN: 4/7
} \\
\hline

\centered{
\,
$\begin{bmatrix}
1 & x & y & 1+xy & \bar{x}y\\
1 & 1 & y & 1    & 1+\bar{x}y\\
1 & 0 & 1 & xy   & 1\\
1 & x & 0 & xy   & \bar{x}y\\
0 & 1 & 0 & 1    & 0\\
0 & 0 & 1 & 0    & 1\\
0 & x & 0 & xy   & 0 \\
0 & 0 & y & 0    & \bar{x}y
\end{bmatrix}$
\,}
& \centered{\, \includegraphics[width=3cm]{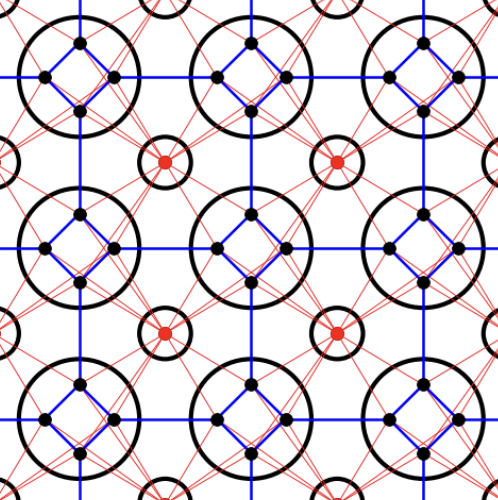} \,} 
& \centered{\, \includegraphics[width=3cm]{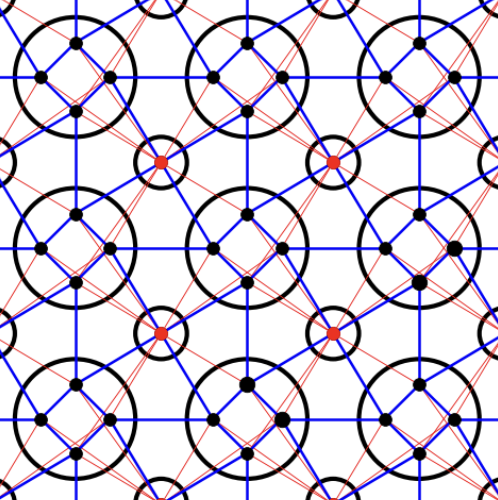} \,} & 
\centered{
type: d, e \\
distance: 4 \\
stabilizers: \{8, 6, 8, 6\} \\
NN hoppings: \{6, 4, 6, 4\} \\
NNN hoppings: \{6, 4, 6, 4\} \\
den-den: 8 \\
planar NN/NNN: yes/no \\
max. conn. NN/NNN: 3/4
} \\
\hline

\centered{
\,
$\begin{bmatrix}
1+\bar{y} & \bar{y} & 1+y & 1+\bar{y} & \bar{y} & x & 1+\bar{y} & xy+\bar{y} \\
1         & 1       & y   & 1         & 0       & 0 & 0         & xy \\
1         & 0       & 1   & 1         & 0       & 0 & 0         & 0 \\
0         & 1       & 0   & 1+y       & 1       & 0 & y         & 0 \\
0         & 0       & 0   & y         & 1       & 1 & y         & 1 \\
0         & 0       & 0   & 0         & 1       & 0 & 1         & 1 \\
0         & 0       & 1   & 0         & 0       & 0 & 0         & 0 \\
0         & 0       & 0   & 0         & 0       & x & 0         & xy \\
0         & 1       & 0   & 1         & 0       & 0 & 0         & 0 \\
0         & 0       & 0   & 0         & 0       & 0 & 1         & 0 \\
0         & 1       & y   & y         & 0       & 0 & y         & 0 \\
0         & 0       & 0   & 0         & 0       & 1 & 0         & 1 
\end{bmatrix}$
\,}
& \centered{\, \includegraphics[width=3cm]{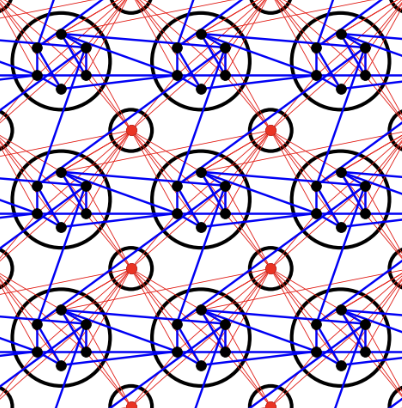} \,} 
& \centered{\, \includegraphics[width=3cm]{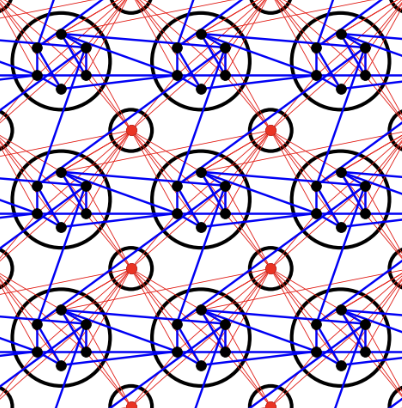} \,} & 
\centered{
type: c, g \\
distance: 4 \\
stabilizers: \{4, 8, 6, 8\} \\
h$_{\text{NN}}$: \{4, 4, 6, 4, 4, 6, 6, 4\} \\
h$_{\text{NNN}}$: \{4, 6, 10, 8, 4, 6, 8, 8\} \\
den-den: 12 \\
planar NN/NNN: yes/yes \\
max. conn. NN/NNN: 6/6
} \\
\hline

\centered{
\,
$\begin{bmatrix}
1 & 0 & 1+y & 0 & x & x & x   & x \\
1 & 1 & 0   & 0 & 0 & x & 0   & xy \\
1 & 0 & y   & 1 & 0 & x & 0   & xy \\
1 & 1 & 1   & 1 & 1 & 0 & 1+y & 0 \\
0 & 1 & 0   & y & 1 & 1 & 0   & 0 \\
0 & 0 & 0   & y & 1 & 0 & y   & 1 \\
0 & 0 & 0   & 0 & 0 & x & 0   & 0 \\
0 & 0 & y   & 0 & 0 & x & 0   & xy \\
0 & 0 & 1   & 1 & 0 & 0 & 0   & 0 \\
0 & 1 & 0   & 0 & 0 & 0 & 0   & 0 \\
0 & 1 & 0   & y & 0 & 0 & y   & 0 \\
0 & 1 & 0   & 0 & 0 & 0 & 1   & 1
\end{bmatrix}$
\,}
& \centered{\, \includegraphics[width=3cm]{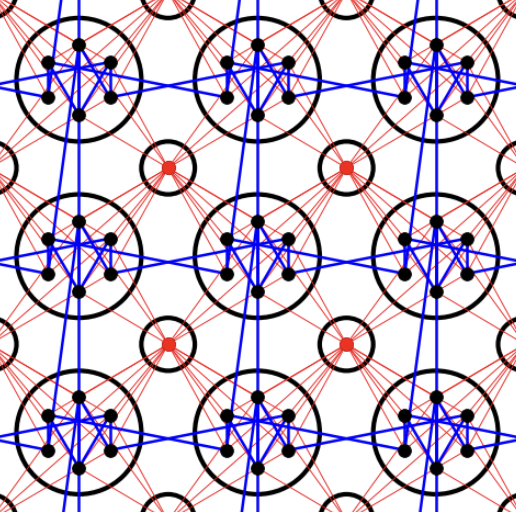} \,} 
& \centered{\, \includegraphics[width=3cm]{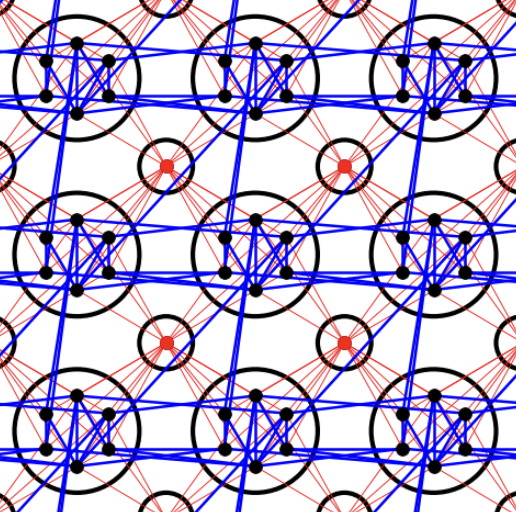} \,} & 
\centered{
type: c, g \\
distance: 4 \\
stabilizers: \{7, 7, 7, 7\} \\
h$_{\text{NN}}$: \{5, 5, 5, 5, 5, 5, 5, 5\} \\
h$_{\text{NNN}}$: \{5, 5, 10, 8, 5, 5, 10, 8\} \\
den-den: 12 \\
planar NN/NNN: no/no \\
max. conn. NN/NNN: 4/7
} \\
\hline
\centered{
$\begin{bmatrix}
1+x                & x          & x          & 1+x+xy \\
1                  & 1          & 0          & xy \\
1+\bar{y}+x\bar{y} & x+x\bar{y} & y+x\bar{y} & 1+x+xy+\bar{y}+x\bar{y} \\
0                  & xy         & 1+y        & \bar{x}y \\
0                  & x          & y          & xy \\
0                  & 0          & 0          & 1
\end{bmatrix}$
}
& \centered{\, \includegraphics[width=3cm]{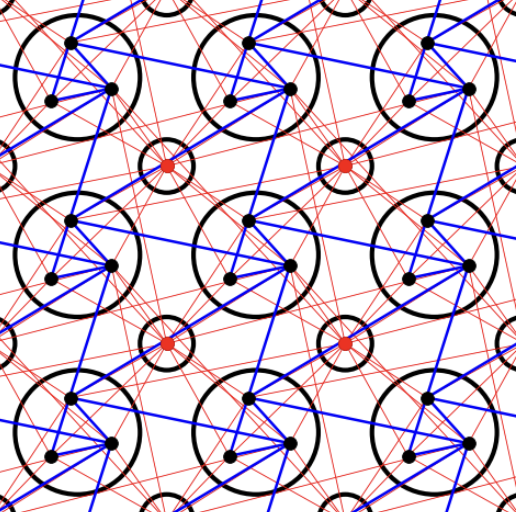} \,} 
& \centered{\, \includegraphics[width=3cm]{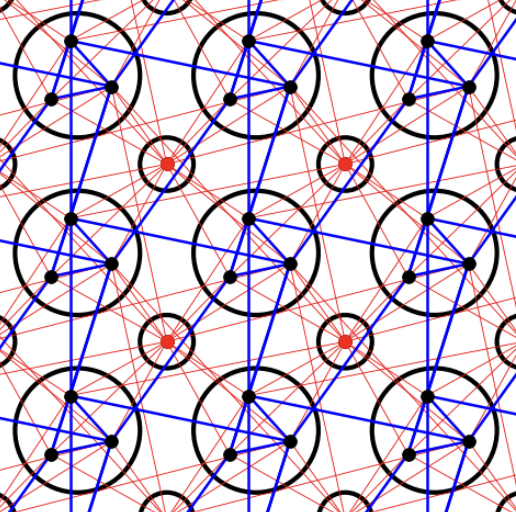} \,} & 
\centered{
type: d, e \\
distance: 5 \\
stabilizers: \{11, 9\} \\
NN hoppings: \{7, 7, 7, 7\} \\
NNN hoppings: \{7, 9, 8, 8\} \\
den-den: 12 \\
planar NN/NNN: yes/no \\
max. conn. NN/NNN: 5/6
} \\
\hline

\end{tabular}
%   \tabfnt{Note: Bold categories are biologically correct growth models.}
\label{table:enc2}
\end{table}

\begin{table}
% \centering
[ht] \caption{} \label{tab:encodings}
% \begin{tabular}{|c|c|c|c|} 
\begin{tabular}{|@{}c@{}|@{}c@{}|@{}c@{}|@{}c@{}|}
\hline Logical operators & NN connectivity & NNN connectivity & Additional information \\
\hline 
\centered{\!\!
$\begin{bmatrix}
1+\bar{y}+x\bar{y} & x+x\bar{y} & 1+x+y+\bar{y}+x\bar{y} & x+xy+\bar{y} \\
1                  & 1          & y                      & 1+xy \\
1                  & 0          & 1+y+\bar{x}y           & 1 \\
0                  & 0          & 1                      & 0 \\
0                  & x          & y                      & xy \\
0                  & 1          & y+\bar{x}y             & 1
\end{bmatrix}$\!\!
}
& \centered{ \includegraphics[width=3cm]{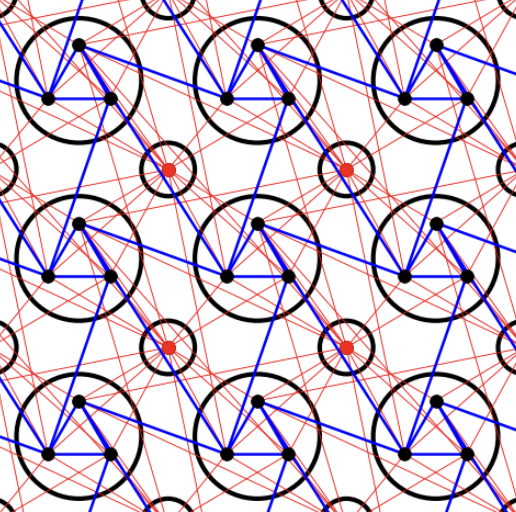} } 
& \centered{ \includegraphics[width=3cm]{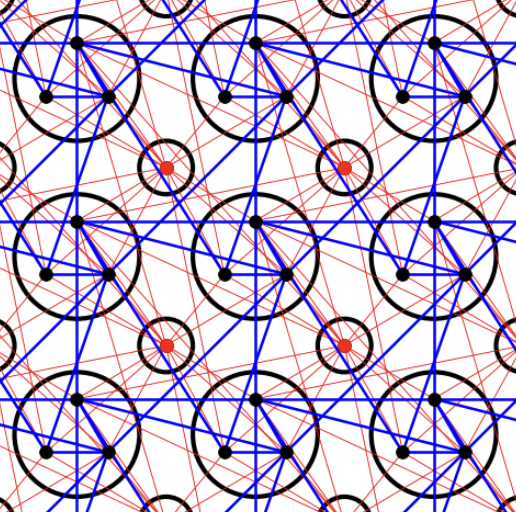} } & 
\centered{
type: c, e \\
distance: 5 \\
stabilizers: \{9, 9\} \\
NN hoppings: \{5, 5, 7, 7\} \\
NNN hoppings: \{6, 6, 7, 9\} \\
den-den: 10 \\
planar NN/NNN: yes/no \\
max. conn. NN/NNN: 5/7
} \\
\hline

\centered{
$\begin{bmatrix}
1       & 0       & 1 & 0   & x & 0 & x & x+xy \\
1       & 1       & 0 & 0   & 0 & 0 & 0 & xy \\
1       & 0       & y & 1+y & 1 & 1 & 0 & xy \\
1       & 0       & 1 & 1+y & 1 & 0 & 1 & 0 \\
0       & 0       & 0 & y   & 1 & 1 & 0 & 0 \\
\bar{x} & \bar{x} & 0 & y   & 1 & 0 & y & 1+y \\
0       & 0       & 0 & 0   & 0 & x & 0 & 0 \\
0       & x       & y & y   & 0 & x & x & x+xy \\
0       & 0       & 1 & 1   & 0 & x & 0 & 0 \\
0       & 1       & 0 & 0   & 0 & 0 & 0 & 0 \\
0       & 1       & 1 & 1+y & 0 & x & y & 0 \\
0       & 1       & 0 & 0   & 0 & 0 & 1 & 1
\end{bmatrix}$
}
& \centered{ \includegraphics[width=3cm]{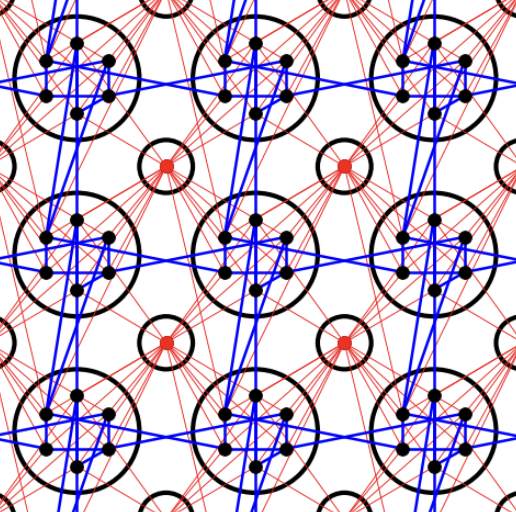} } 
& \centered{ \includegraphics[width=3cm]{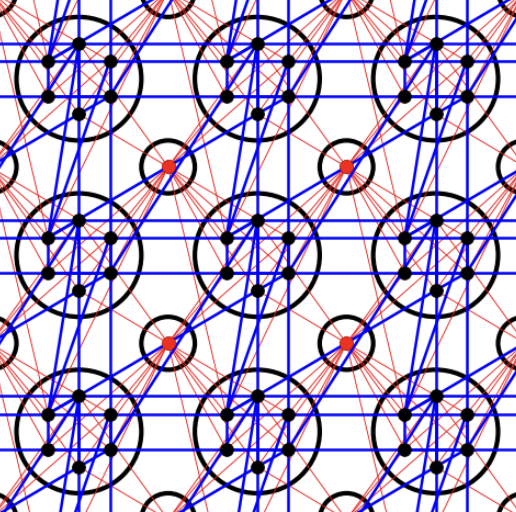} } & 
\centered{
type: c, g \\
distance: 5 \\
stabilizers: \{8, 8, 8, 8\} \\
h$_{\text{NN}}$: \{6, 8, 6, 8, 6, 8, 6, 8\} \\
h$_{\text{NNN}}$: \{9, 5, 11, 11, 9, 5, 11, 11\} \\
den-den: 12 \\
planar NN/NNN: no/no \\
max. conn. NN/NNN: 4/7
} \\
\hline

\centered{\!\!
$\begin{bmatrix}
1+\bar{x} & x & y+\bar{x} & 1+x+y & \bar{x}+\bar{x}y \\
1         & 1 & y         & 1     & 1+\bar{x}y \\
1         & 1 & 1         & 1+xy  & 1 \\
1         & x & 0         & xy    & \bar{x}y \\
0         & 1 & 0         & 1     & 0 \\
0         & 0 & 1         & 0     & 1 \\
0         & 0 & 0         & x+xy  & 0 \\
0         & 0 & y         & 0     & \bar{x}y
\end{bmatrix}$\!\!
}
& \centered{ \includegraphics[width=3cm]{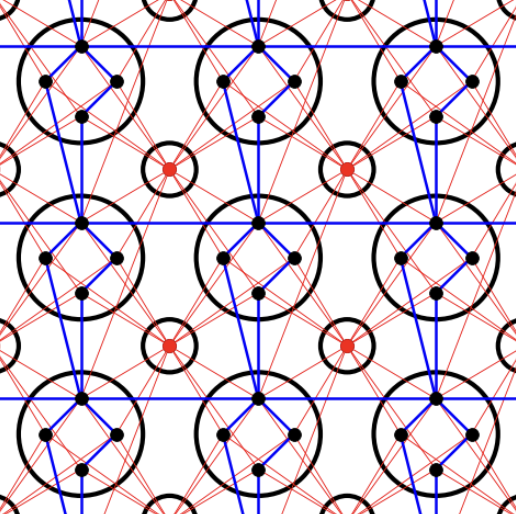} } 
& \centered{ \includegraphics[width=3cm]{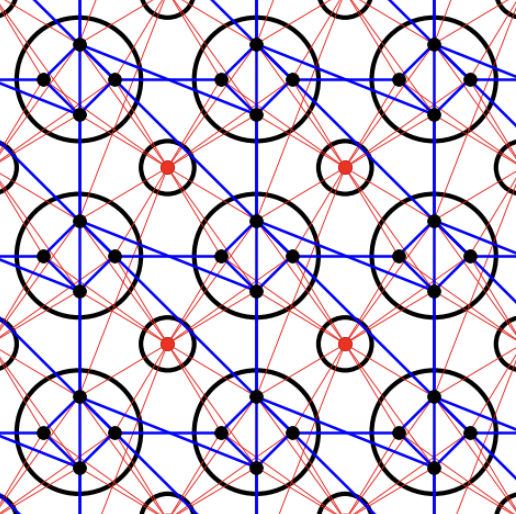} } & 
\centered{
type: d, e \\
distance: 5 \\
stabilizers: \{10, 6, 10 8\} \\
NN hoppings: \{5, 5, 6, 6\} \\
NNN hoppings: \{8, 8, 6, 5\} \\
den-den: 10 \\
planar NN/NNN: yes/no \\
max. conn. NN/NNN: 6/6
} \\
\hline

\centered{\!\!
$\begin{bmatrix}
1+x+\bar{y} & x+xy+\bar{y}+x\bar{y} & x+y & 1+x+xy+y \\
1           & 1                     & 0   & xy \\
1           & x                     & y   & 1+x+xy \\
1           & 1+x                   & 0   & xy \\
1+x         & 0                     & x+y & 1+x \\
0           & x                     & 1   & 1
\end{bmatrix}$\!\!
}
& \centered{ \includegraphics[width=3cm]{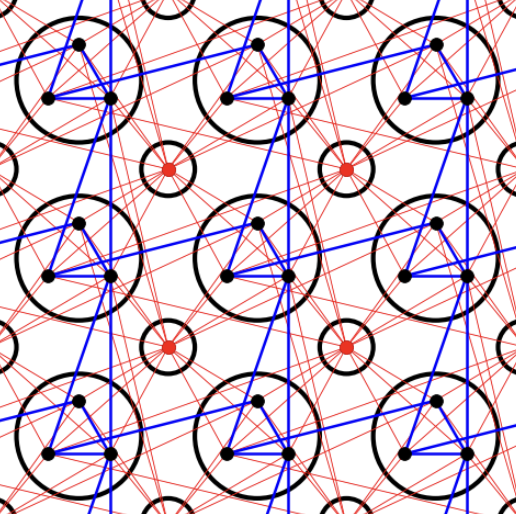} } 
& \centered{ \includegraphics[width=3cm]{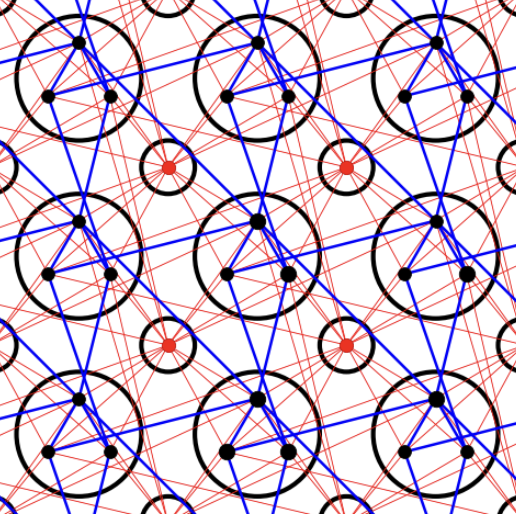} } & 
\centered{
type: c, e \\
distance: 6 \\
stabilizers: \{12, 8\} \\
NN hoppings: \{8, 8, 7, 7\} \\
NNN hoppings: \{9, 9, 11, 9\} \\
den-den: 12 \\
planar NN/NNN: yes/no \\
max. conn. NN/NNN: 5/6
} \\
\hline

\centered{\!\!
$\begin{bmatrix}
1+\bar{x}+\bar{y} & x+\bar{y} & 1+y+\bar{x} & 1+xy+\bar{y} & \bar{x}y+\bar{x}+\bar{y} \\
1                 & 1         & y           & 1            & 1+\bar{x}y \\
1                 & 0         & 1+xy        & xy           & 1+y \\
1                 & 1+x       & y           & xy           & \bar{x} \\
0                 & 1         & 0           & 1            & \bar{x}y \\
0                 & y         & 1           & y            & 1 \\
0                 & x         & xy          & xy           & y \\
0                 & 0         & y           & 0            & 0
\end{bmatrix}$\!\!
}
& \centered{ \includegraphics[width=3cm]{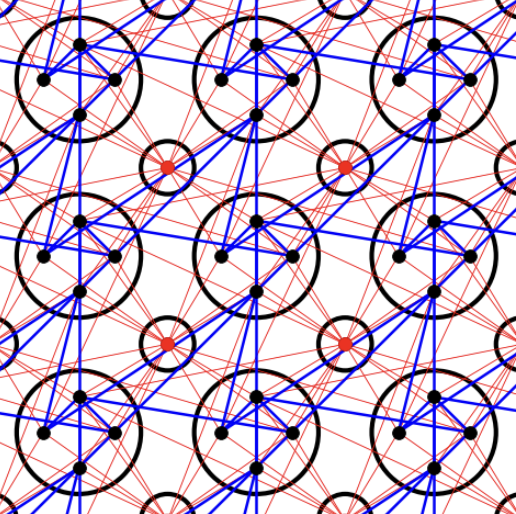} } 
& \centered{ \includegraphics[width=3cm]{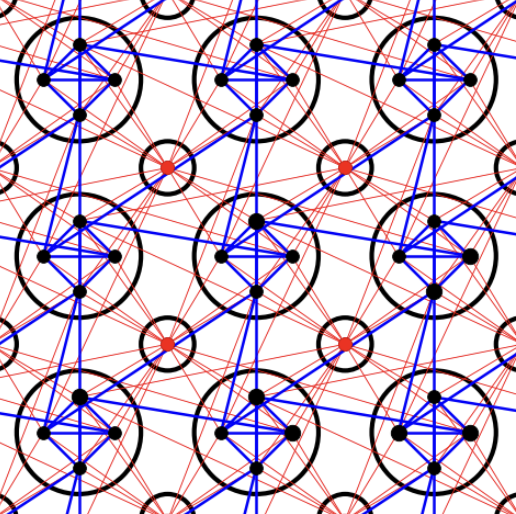} } & 
\centered{
type: d, e \\
distance: 6 \\
stabilizers: \{8, 10, 12, 10\} \\
NN hoppings: \{7, 7, 7, 7\} \\
NNN hoppings: \{8, 8, 7, 7\} \\
den-den: 12 \\
planar NN/NNN: no/no \\
max. conn. NN/NNN: 5/6
} \\
\hline

\centered{\!\!
$\begin{bmatrix}
1+x+\bar{x} & x+\bar{x} & 1+\bar{x} & xy+\bar{x} & x+y+\bar{x}y+\bar{x} \\
1           & 1         & 0         & 1          & 1+\bar{x}y \\
1           & x         & x+xy      & x+xy       & 1 \\
1           & x         & xy        & xy         & \bar{x}y \\
0           & 1         & 0         & 1          & 0 \\
0           & \bar{x}   & 1         & \bar{x}    & 1 \\
0           & 0         & x+xy      & x+xy       & 0 \\
0           & 0         & 0         & 0          & \bar{x}y
\end{bmatrix}$\!\!
}
& \centered{ \includegraphics[width=3cm]{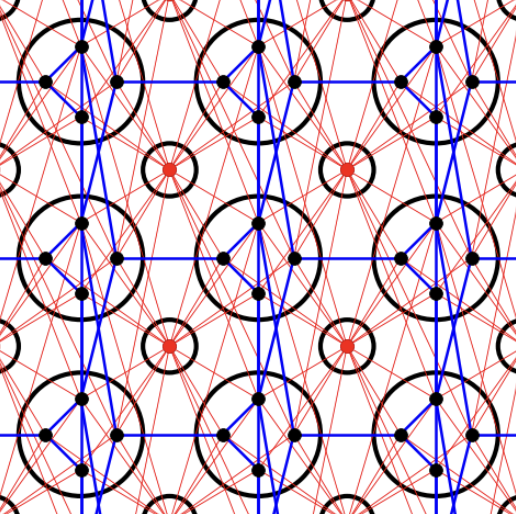} } 
& \centered{ \includegraphics[width=3cm]{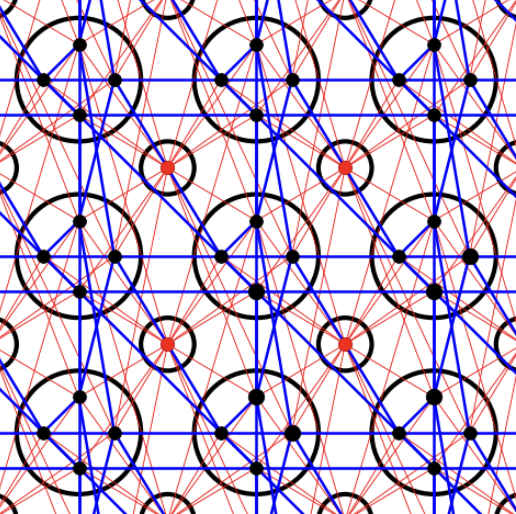} } & 
\centered{
type: d, e \\
distance: 6 \\
stabilizers: \{9, 11, 9, 11\} \\
NN hoppings: \{6, 6, 8, 6\} \\
NNN hoppings: \{9, 9, 7, 6\} \\
den-den: 12 \\
planar NN/NNN: no/no \\
max. conn. NN/NNN: 5/6
} \\
\hline

\centered{\!\!
$\begin{bmatrix}
1         & x   & y           & 1+xy & \bar{x}y \\
\bar{x}   & x   & \bar{x}     & 1+xy & 1+\bar{x}+\bar{x}y \\
1         & 0   & 1+xy        & xy   & 1+y \\
1+\bar{x} & 1+x & y+\bar{x}   & xy   & \bar{x} \\
0         & 1   & 0           & 1    & 0 \\
1+\bar{x} & 1+x & 1+y+\bar{x} & xy   & 1+\bar{x} \\
0         & 0   & xy          & xy   & y \\
\bar{x}   & x   & 1+y+\bar{x} & 1+xy & \bar{x}
\end{bmatrix}$\!\!
}
& \centered{ \includegraphics[width=3cm]{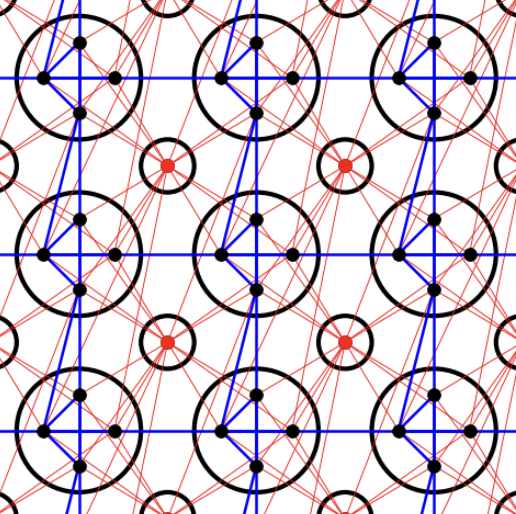} } 
& \centered{ \includegraphics[width=3cm]{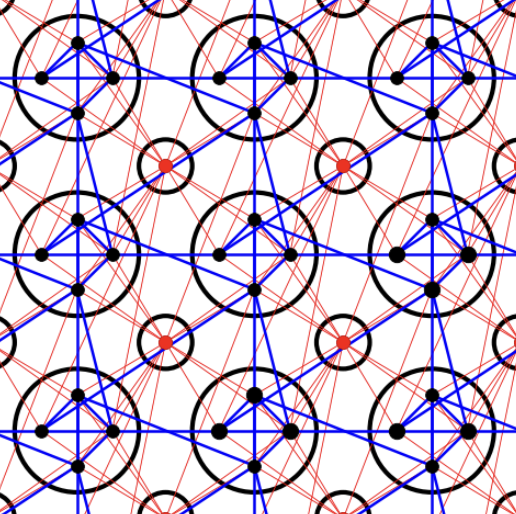} } & 
\centered{
type: d, e \\
distance: 6 \\
stabilizers: \{8, 8, 10, 8\} \\
NN hoppings: \{8, 6, 6, 10\} \\
NNN hoppings: \{10, 8, 6, 8\} \\
den-den: 12 \\
planar NN/NNN: no/no \\
max. conn. NN/NNN: 6/6
} \\
\hline

\end{tabular}
%   \tabfnt{Note: Bold categories are biologically correct growth models.}
\label{table:enc3}
\end{table}

\begin{table}
% \centering
[ht] \caption{} \label{tab:encodings}
% \begin{tabular}{|c|c|c|c|} 
\begin{tabular}{|@{}c@{}|@{}c@{}|@{}c@{}|@{}c@{}|}
\hline Logical operators & NN connectivity & NNN connectivity & Additional information \\
\hline 
\centered{\!\!
$\begin{bmatrix}
x+\bar{y}+x\bar{y} & x\bar{y}   & x+y+x\bar{y} & \bar{y}+x\bar{y} \\
1                  & 1          & 0            & xy \\
1+\bar{y}+x\bar{y} & x+x\bar{y} & y+x\bar{y}   & 1+x+xy+\bar{y}+x\bar{y} \\
0                  & x+xy       & y            & y \\
0                  & x          & y            & xy \\
0                  & xy         & 1+y          & 1+y
\end{bmatrix}$\!\!
}
& \centered{ \includegraphics[width=3cm]{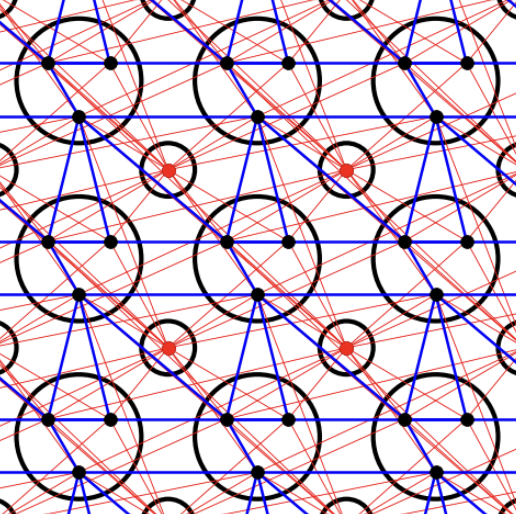} } 
& \centered{ \includegraphics[width=3cm]{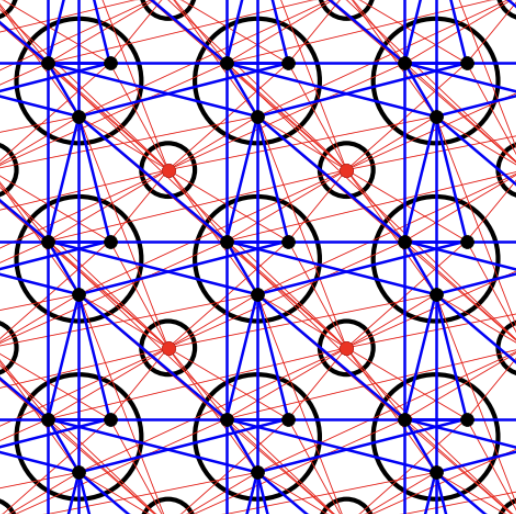} } & 
\centered{
type: c, e \\
distance: 7 \\
stabilizers: \{14, 10\} \\
NN hoppings: \{8, 8, 7, 9\} \\
NNN hoppings: \{8, 12, 10, 10\} \\
den-den: 14 \\
planar NN/NNN: yes/no \\
max. conn. NN/NNN: 6/9
} \\
\hline

\centered{\!\!
$\begin{bmatrix}
1+\bar{x}y        & x         & y+\bar{x}y       & x+xy+\bar{x}y \\
1+xy              & 1         & xy+y             & 1 \\
1+\bar{x}+\bar{y} & x\bar{y}  & \bar{x}          & 1+x+\bar{x}+x\bar{y} \\
0                 & 1+y       & 1+x+\bar{x}y     & 0 \\
0                 & x         & y                & xy \\
0                 & 1+\bar{y} & \bar{x}+x\bar{y} & 0
\end{bmatrix}$\!\!
}
& \centered{ \includegraphics[width=3cm]{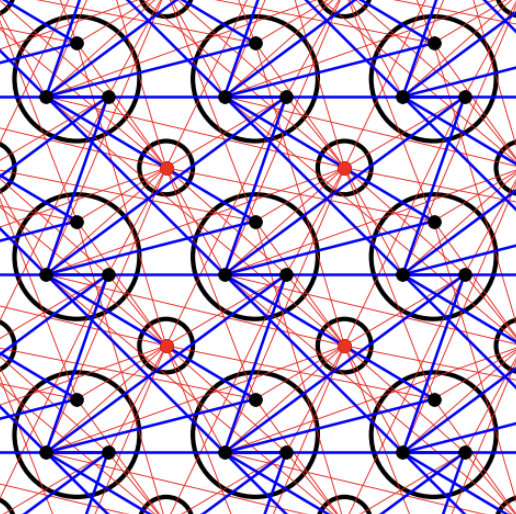} } 
& \centered{ \includegraphics[width=3cm]{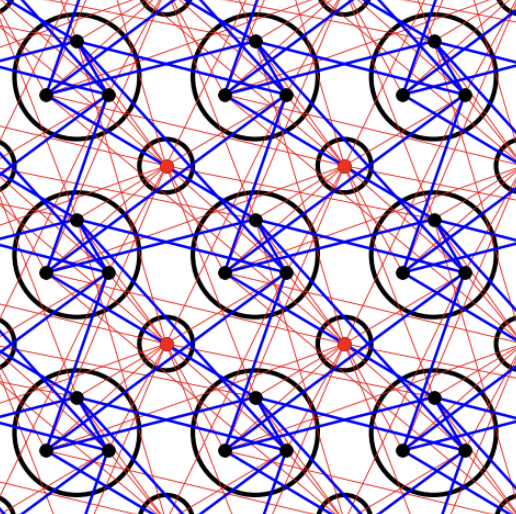} } & 
\centered{
type: c, e \\
distance: 7 \\
stabilizers: \{12, 12\} \\
NN hoppings: \{10, 8, 10, 12\} \\
NNN hoppings: \{7, 11, 9, 11\} \\
den-den: 14 \\
planar NN/NNN: yes/no \\
max. conn. NN/NNN: 7/5
} \\
\hline

\end{tabular}
%   \tabfnt{Note: Bold categories are biologically correct growth models.}
\label{table:enc4}
\end{table}

% $\begin{bmatrix}
% 0 & 0 & 0 & 0\\
% 0 & 0 & 0 & 0\\
% 0 & 0 & 0 & 0\\
% 0 & 0 & 0 & 0\\
% 0 & 0 & 0 & 0\\
% 0 & 0 & 0 & 0
% \end{bmatrix}$

% $\begin{bmatrix}
% 0 & 0 & 0 & 0\\
% 0 & 0 & 0 & 0\\
% 0 & 0 & 0 & 0\\
% 0 & 0 & 0 & 0\\
% 0 & 0 & 0 & 0\\
% 0 & 0 & 0 & 0
% \end{bmatrix}$

\end{document}